\documentclass[11pt,a4paper]{article}
\pdfoutput=1

\usepackage{jheppub}

\usepackage[utf8]{inputenc}
\usepackage[T1]{fontenc}
\usepackage{afterpage}

\newcommand{\Dmq}{\Delta m^2}
\newcommand{\Eps}{\varepsilon}
\newcommand{\Epx}{\mathcal{E}}
\newcommand{\diag}{\mathop{\mathrm{diag}}}
\newcommand{\Nuc}[2]{{\ensuremath{\mbox{}^{#1}}\text{#2}}}
\newenvironment{pagefigure}{\begin{figure}[!p]}{\afterpage\clearpage\end{figure}}

\makeatletter
\DeclareRobustCommand\recite[1]{\begingroup\@fileswfalse\cite{#1}\endgroup}
\gdef\@fpheader{} 
\makeatother

\allowdisplaybreaks

\title{Neutrino Oscillation Constraints on $U(1)'$ Models: from
  Non-Standard Interactions to Long-Range Forces}

\author[a]{Pilar Coloma,}
\affiliation[a]{Instituto de Física Teórica UAM/CSIC, Calle de
  Nicolás Cabrera 13--15, Universidad Autónoma de Madrid,
  Cantoblanco, E-28049 Madrid, Spain}
\emailAdd{pilar.coloma@ift.csic.es}

\author[b,c,d]{M.~C.~Gonzalez-Garcia,}
\affiliation[b]{Departament de Física Quàntica i Astrofísica and
  Institut de Ciències del Cosmos, Universitat de Barcelona, Diagonal
  647, E-08028 Barcelona, Spain}
\affiliation[c]{Institució Catalana de Recerca i Estudis Avançats
  (ICREA), Pg.\ Lluis Companys 23, E-08010 Barcelona, Spain}
\affiliation[d]{C.N.~Yang Institute for Theoretical Physics, Stony
  Brook University, Stony Brook, NY 11794-3840, USA}
\emailAdd{maria.gonzalez-garcia@stonybrook.edu}

\author[a]{Michele Maltoni}
\emailAdd{michele.maltoni@csic.es}

\abstract{We quantify the effect of gauge bosons from a weakly coupled
  lepton flavor dependent $U(1)'$ interaction on the matter background
  in the evolution of solar, atmospheric, reactor and long-baseline
  accelerator neutrinos in the global analysis of oscillation data.
  The analysis is performed for interaction lengths ranging from the
  Sun-Earth distance to effective contact neutrino interactions. We
  survey $\sim 10000$ set of models characterized by the six relevant
  fermion $U(1)'$ charges and find that in all cases, constraints on
  the coupling and mass of the $Z'$ can be derived.  We also find that
  about 5\% of the $U(1)'$ model charges lead to a viable LMA-D
  solution but this is only possible in the contact interaction
  limit. We explicitly quantify the constraints for a variety of
  models including $U(1)_{B-3L_e}$, $U(1)_{B-3L_\mu}$,
  $U(1)_{B-3L_\tau}$, $U(1)_{B-\frac{3}{2}(L_\mu+L_\tau)}$,
  $U(1)_{L_e-L_\mu}$, $U(1)_{L_e-L_\tau}$,
  $U(1)_{L_e-\frac{1}{2}(L_\mu+L_\tau)}$.  We compare the constraints
  imposed by our oscillation analysis with the strongest bounds from
  fifth force searches, violation of equivalence principle as well as
  bounds from scattering experiments and white dwarf cooling.  Our
  results show that generically, the oscillation analysis improves
  over the existing bounds from gravity tests for $Z'$ lighter than
  $\sim 10^{-11}$~eV.  In the contact interaction limit, we find that
  for most models listed above there are values of $g'$ and $M_{Z'}$
  for which the oscillation analysis provides constraints beyond those
  imposed by laboratory experiments. Finally we illustrate the range
  of $Z'$ and couplings leading to a viable LMA-D solution for two
  sets of models.}

\preprint{IFT-UAM/CSIC-20-133, YITP-SB-2020-30}


\begin{document}

\maketitle

\section{Introduction}

Experiments measuring the flavor composition of neutrinos produced in
the Sun, in the Earth's atmosphere, in nuclear reactors and in
particle accelerators have established that lepton flavor is not
conserved in neutrino propagation, but it oscillates with a wavelength
which depends on distance and energy. This demonstrates beyond doubt
that neutrinos are massive and that the mass states are non-trivial
admixtures of flavor states~\cite{Pontecorvo:1967fh, Gribov:1968kq},
see Ref.~\cite{GonzalezGarcia:2007ib} for an overview.

When traveling through matter, the flavor evolution of the neutrino
ensemble is affected by the difference in the effective potential
induced by elastic forward scattering of neutrino with matter, the
so-called Mikheev-Smirnov-Wolfenstein (MSW)
mechanism~\cite{Wolfenstein:1977ue, Mikheev:1986gs}.  Within the
context of the Standard Model (SM) of particle interactions, this
effect is fully determined and leads to a matter potential which, for
neutral matter, is proportional to the number density of electrons at
the neutrino position, $V=\sqrt{2} G_F N_e(r)$, and which only affects
electron neutrinos.  New flavor dependent interactions can modify the
matter potential and consequently alter the pattern of flavor
transitions, thus leaving imprints in the oscillation data involving
neutrinos which have traveled through large regions of matter, as is
the case for solar and atmospheric neutrinos.

Forward elastic scattering takes place in the limit of zero momentum
transfer, so as long as the range of the interaction is shorter than
the scale over which the matter density extends, the effective matter
potential can be obtained in the contact interaction approximation
between the neutrinos and the matter particles. The paradigmatic
example is provided by neutral current non-standard interactions
(NSI)~\cite{Wolfenstein:1977ue, Valle:1987gv, Guzzo:1991hi} between
neutrinos and matter (for recent reviews, see~\cite{Ohlsson:2012kf,
  Miranda:2015dra, Farzan:2017xzy, Dev:2019anc}), which can be
parametrized as
\begin{equation}
  \label{eq:NSILagrangian}
  \mathcal L_\text{NSI} = -2\sqrt2 G_F
  \sum_{f,P,\alpha,\beta} \Eps_{\alpha\beta}^{f,P}
  (\bar\nu_\alpha\gamma^\mu P_L\nu_\beta)
  (\bar f\gamma_\mu P f) \,,
\end{equation}
where $G_F$ is the Fermi constant, $\alpha, \beta$ are flavor indices,
$P\equiv P_L, P_R$ and $f$ is a SM charged fermion.  These operators
are expected to arise generically from the exchange of some mediator
state heavy enough for the contact interaction approximation to hold.
In this notation, $\Eps_{\alpha\beta}^{f,P}$ parametrizes the strength
of the new interaction with respect to the Fermi constant,
$\Eps_{\alpha\beta}^{f,P} \sim \mathcal{O}(G_X/G_F)$. Generically they
modify the matter potential in neutrino propagation, but --~being
local interactions~-- the resulting potential is still proportional to
the number density of particles in the medium at the neutrino
position.  Since such modifications arise from a coherent effect,
oscillation bounds apply even to NSI induced by ultra light mediators,
as long as their interaction length is shorter than the neutrino
oscillation length. For the experiments considered here, such
condition is fulfilled as long as $M_{Z'}\gtrsim
10^{-12}$~eV~\cite{GonzalezGarcia:2006vp}.

Conversely, if the mediator is too light then the contact interaction
approximation is no longer valid, and the flavor dependent forces
between neutrino and matter particles become long-range. In this case
neutrino propagation can still be described in terms of a matter
potential, which however is no longer simply determined by the number
density of particles in the medium at the neutrino position, but it
depends instead on the average of the matter density within a radius
$\sim 1/M_{Z'}$ around it~\cite{Grifols:2003gy, Joshipura:2003jh,
  GonzalezGarcia:2006vp, Davoudiasl:2011sz, Wise:2018rnb}.

At present, the global analysis of data from oscillation experiments
provides some of the strongest constraints on the size of the NSI
affecting neutrino propagation~\cite{GonzalezGarcia:2011my,
  Gonzalez-Garcia:2013usa, Esteban:2018ppq}. Analysis of early
oscillation data was also used to impose constraints on flavor
dependent long-range forces~\cite{Grifols:2003gy, Joshipura:2003jh,
  GonzalezGarcia:2006vp}.

Straightforward constructions leading to Eq.~\eqref{eq:NSILagrangian}
have an extended gauge sector with an additional $U(1)'$ symmetry with
charge involving some of the lepton flavors and an heavy enough gauge
boson.  Conversely if the gauge boson is light enough a long range
force will be generated. Thus the analysis of neutrino oscillation
data can shed light on the valid range of $Z'$ mass and coupling in
both regimes. Following this approach, the marginalized bounds on the
NSI coefficients derived from the global analysis of oscillations in
presence of NSI performed in Ref.~\cite{Esteban:2018ppq} were adapted
to place constraints on the coupling and mass of the new gauge boson
both in the NSI limit~\cite{Heeck:2018nzc} and in the long-range
regime~\cite{Wise:2018rnb} for several $U(1)'$ flavor symmetries.
However, strictly speaking, the bounds derived in
Ref.~\cite{Esteban:2018ppq} cannot be directly used to constraint the
$U(1)'$ scenarios because in the latter case only flavor diagonal
interactions (and only some of them depending on the $U(1)'$ charge)
are generated, while the bounds derived in Ref.~\cite{Esteban:2018ppq}
were obtained in the most general parameter space with all relevant
four-fermion interactions (flavor conserving and flavor changing)
being simultaneously non-vanishing. In order to derive statistically
consistent bounds on each $U(1)'$ scenario a dedicated analysis has to
be performed in its reduced parameter space.

With this motivation, in this work we perform such dedicated global
analysis of oscillation data in the framework of lepton flavor
dependent $U(1)'$ interactions which affect the neutrino evolution in
matter, with interaction lengths ranging from the Sun-Earth distance
to effective contact neutrino interactions. In Sec.~\ref{sec:forma} we
describe the models which will be studied and derive the matter
potential generated both in the contact interaction limit (in
Sec.~\ref{sec:NSI}) and in the case of finite interaction range (in
Sec.~\ref{sec:LR}) as a function of the $U(1)'$ charges.  The results
of the global analysis are presented in Sec.~\ref{sec:results}.  In
particular the bounds imposed by the analysis and how they compare
with those from other experiments are presented in
Sec.~\ref{sec:bounds}.  An additional consideration that we take into
account is that in the presence of NSI a degeneracy exists in
oscillation data, leading to the so called LMA-Dark
(LMA-D)~\cite{Miranda:2004nb} solution first observed in solar
neutrinos, where for suitable NSI the data can be explained by a
mixing angle $\theta_{12}$ in the second octant. For this new solution
to appear the new interactions must be such that the matter potential
difference for electron neutrinos reverses its sign with respect to
that in the SM. It is not trivial to generate such large effects
without conflicting with bounds from other experiments, though models
with light mediators (\textit{i.e.}, below the electroweak scale) have
been proposed as viable candidates~\cite{Farzan:2017xzy,
  Miranda:2015dra, Farzan:2015doa, Farzan:2015hkd, Babu:2017olk,
  Denton:2018xmq}.  Section~\ref{sec:dark} contains our findings on
viable models for LMA-D. We summarize our conclusions in
Sec.~\ref{sec:conclu}.  We present some details of the translation of
the bounds from some experiments to the models studied in an appendix.

\section{Formalism}
\label{sec:forma}

We are going to focus on $U(1)'$ models which can be tested in
neutrino oscillation experiments via its contribution to the matter
potential. As a start this requires that the new gauge boson couples
to the fermions of the first generation.

An important issue when enlarging the Standard Model with a new
$U(1)'$ gauge group is the possibility of mixing between the three
neutral gauge bosons of the model which can, in general, be induced in
either kinetic or mass terms. While kinetic mixing is fairly generic
as it can be generated at the loop level with the SM particle
contents, matter mixing is model dependent as it requires an extended
scalar sector with a vacuum expectation value charged both under the
SM and the $U(1)'$.  In what respects the effect of the new $U(1)'$ in
oscillation experiments an important observation is that if the new
interaction does not couple directly to fermions of the first
generation, no matter effects can be generated by kinetic
mixing~\cite{Heeck:2018nzc}. Thus neglecting mixing effects yields the
most model independent and conservative bounds from neutrino
oscillation results.  So in what follows we are going to work under
the assumption that the $Z'$ mixing with SM gauge bosons can be safely
neglected.

In addition we notice that only vector interactions contribute to the
matter potential in neutrino propagation.  Altogether the part of the
$U(1)'$ Lagrangian relevant for propagation in ordinary matter has the
most general form
\begin{multline}
\label{eq:zpcurrent}
  \mathcal{L}_{Z'}^\text{matter}
  = -g'\big( a_u\, \bar{u}\gamma^\alpha u
  + a_d\, \bar{d}\gamma^\alpha d
  + a_e\, \bar{e}\gamma^\alpha e
  \\
  + b_e \,\bar{\nu}_e\gamma^\alpha P_L \nu_e
  + b_\mu\,\bar{\nu}_\mu\gamma^\alpha P_L \nu_\mu
  + b_\tau\, \bar{\nu}_\tau \gamma^\alpha P_L \nu_\tau \big)  Z'_\alpha
\end{multline}
with arbitrary charges $a_{u,d,e}$ and $b_{e,\mu,\tau}$.

These charges can be accommodated in generalized anomaly free
UV-complete models including only the SM particles plus right-handed
neutrinos~\cite{Allanach:2018vjg}.  If, in addition, one requires all
couplings to be vector-like and the quark couplings to be generation
independent, the condition of anomaly cancellation for models with
only SM plus right handed neutrinos imposes constrains over the six
charges above and one ends with a subclass of models characterized by
three independent charges which can be chosen to be, for example,
$B-L$, $(L_\mu-L_\tau)$, and $(L_\mu-L_e)$~\cite{Heeck:2018nzc}:
\begin{equation}
  \label{eq:anomod}
  c_\textsc{bl} (B-L) + c_{\mu\tau} (L_\mu-L_\tau)+ c_{\mu e} (L_\mu-L_e) \,,
\end{equation}
so $a_u = a_d = c_\textsc{bl} / 3$, $a_e = b_e = -(c_\textsc{bl} +
c_{\mu e})$, $b_\mu = -c_\textsc{bl} + c_{\mu e} + c_{\mu\tau}$, and
$b_\tau = -(c_\textsc{bl} + c_{\mu\tau})$.  In particular models with
charges as in Eq.~\eqref{eq:anomod} are not constrained by rare $Z$
decays or flavor changing neutral current (FCNC) meson
decays~\cite{Dror:2017ehi, Dror:2017nsg, Dror:2018wfl}.  For
convenience, in table~\ref{tab:models} we list the charges
corresponding to some of the models discussed in the literature.
\begin{table}\centering
  \catcode`!=\active\def!{\hphantom{-}}
  \catcode`?=\active\def?{\hphantom{0}}
  \begin{tabular}{@{\rule[-2.1mm]{0pt}{6mm}}|c|cccccc|}
    \hline
    Model
    & $a_u$ & $a_d$ & $a_e$ & $b_e$ & $b_\mu$ & $b_\tau$
    \\
    \hline
    $B - 3L_e$
    & $\frac{1}{3}$ & $\frac{1}{3}$ & $-3$ & $-3$ & $!0$ & $!0$
    \\
    $B - 3L_\mu$
    & $\frac{1}{3}$ & $\frac{1}{3}$ & $!0$ & $!0$ & $-3$ & $!0$
    \\
    $B - 3L_\tau$
    & $\frac{1}{3}$ & $\frac{1}{3}$ & $!0$ & $!0$ & $!0$ & $-3$
     \\
    $B - \frac{3}{2}(L_\mu + L_\tau)$
    & $\frac{1}{3}$ & $\frac{1}{3}$ & $!0$ & $!0$ & $-\frac{3}{2}$ & $-\frac{3}{2}$
    \\
    $L_e - L_\mu$
    & $0$ & $0$ & $!1$ & $!1$ & $-1$ & $!0$
    \\
    $L_e - L_\tau$
    & $0$ & $0$ & $!1$ & $!1$ & $!0$ & $-1$
    \\
    $L_e - \frac{1}{2}(L_\mu + L_\tau)$
    & $0$ & $0$ & $!1$ & $!1$ & $-\frac{1}{2}$ & $-\frac{1}{2}$
    \\
    $B_y + L_\mu + L_\tau$ Ref.~\recite{Farzan:2015doa}
    & $\frac{1}{3}$ & $\frac{1}{3}$ & $!0$ & $!0$ & $!1$ & $!1$
    \\
    $L_e + 2L_\mu + 2L_\tau$
    & $0$ & $0$ & $!1$ & $!1$ & $!2$ & $!2$
    \\
    \hline
  \end{tabular}
  \caption{Relevant charges for the matter effects in neutrino
    oscillation experiments corresponding to a selection of models
    studied in the literature.  For the model presented in
    Ref.~\recite{Farzan:2015doa} we have defined $B_y \equiv B_1 - y
    B_2 - (3-y)B_3$ where $y$ is an arbitrary constant. The results
    presented in this article are independent of $y$.}
  \label{tab:models}
\end{table}

In general, the evolution of the neutrino and antineutrino flavor
state during propagation is governed by the Hamiltonian:
\begin{equation}
  H^\nu = H_\text{vac} + H_\text{mat}
  \quad\text{and}\quad
  H^{\bar\nu} = ( H_\text{vac} - H_\text{mat} )^* \,,
\end{equation}
where $H_\text{vac}$ is the vacuum part which in the flavor basis
$(\nu_e, \nu_\mu, \nu_\tau)$ reads
\begin{equation}
  \label{eq:Hvac}
  H_\text{vac} = U_\text{vac} D_\text{vac} U_\text{vac}^\dagger
  \quad\text{with}\quad
  D_\text{vac} = \frac{1}{2E_\nu} \diag(0, \Dmq_{21}, \Dmq_{31}) \,.
\end{equation}
Here $U_\text{vac}$ denotes the three-lepton mixing matrix in
vacuum~\cite{Pontecorvo:1967fh, Maki:1962mu, Kobayashi:1973fv}.
Following the convention of Ref.~\cite{Coloma:2016gei}, we define
$U_\text{vac} = R_{23}(\theta_{23}) R_{13}(\theta_{13})
\tilde{R}_{12}(\theta_{12}, \delta_\text{CP})$, where
$R_{ij}(\theta_{ij})$ is a rotation of angle $\theta_{ij}$ in the $ij$
plane and $\tilde{R}_{12}(\theta_{12},\delta_\text{CP})$ is a complex
rotation by angle $\theta_{12}$ and phase $\delta_\text{CP}$.

Concerning the matter part $H_\text{mat}$ of the Hamiltonian generated
by the SM together with the $U(1)'$ interactions
in~\eqref{eq:zpcurrent}, its form depends on the new interaction
length determined by the $Z'$ mass as we discuss next.

\subsection{The large ($M_{Z'}\gtrsim 10^{-12}$~eV) $M_{Z'}$ limit: the NSI regime}
\label{sec:NSI}

In the limit of large $M_{Z'}$, the $Z'$ field can be integrated out
from the spectrum and Eq.~\eqref{eq:zpcurrent} generate effective
dimension-six four-fermion interactions leading to Neutral Current NSI
between neutrinos and matter which are usually parametrized in the
form of Eq.~\eqref{eq:NSILagrangian}.  The coefficients
$\Eps_{\alpha\beta}^{f,P}$ parametrizes the strength of the new
interaction with respect to the Fermi constant, with
\begin{equation}
  \Eps_{\alpha\beta}^{f,L}=\Eps_{\alpha\beta}^{f,R}
  = \delta_{\alpha\beta} \,
  \frac{1}{2\sqrt{2} G_F
  } \, \frac{g'^2}{M_{Z'}^2} \,
  a_f\, b_\alpha
  \equiv \delta_{\alpha\beta} \frac{1}{2}\, a_f\, b_\alpha\, \Eps^0
\end{equation}
where we have introduced the notation
\begin{equation}
  \label{eq:ep0}
  \Eps^0 \equiv \frac{1}{\sqrt{2} G_F}\frac{g'^2}{M_{Z'}^2}
\end{equation}
As it is well known, only vector NSI contribute to the matter
potential in neutrino oscillations. It is therefore convenient to
define the parameters relevant for neutrino oscillation experiments
as:
\begin{equation}
  \label{eq:eps-xi}
  \Eps_{\alpha\beta}^f
  \equiv \Eps_{\alpha\beta}^{f,L} + \Eps_{\alpha\beta}^{f,R} =
  \delta_{\alpha\beta} \, a_f \,b_\alpha \,\Eps^0 \,.
\end{equation}
These interactions lead to a flavor diagonal modification of the
matter potential
\begin{equation}
  \label{eq:Hmat}
  H_\text{mat} = \sqrt{2} G_F N_e(\vec{x})
  \begin{pmatrix}
    1+\Epx_{ee}(\vec{x}) & 0 & 0 \\
    0 & \Epx_{\mu\mu}(\vec{x}) & 0 \\
    0 & 0 & \Epx_{\tau\tau}(\vec{x})
  \end{pmatrix}
\end{equation}
where the ``$+1$'' term in the $ee$ entry accounts for the standard
contribution, and
\begin{equation}
  \label{eq:epx-nsi}
  \Epx_{\alpha\alpha}(\vec{x}) = \sum_{f=e,u,d}
  \frac{N_f(\vec{x})}{N_e(\vec{x})} \Eps_{\alpha\alpha}^f
\end{equation}
describes the non-standard part. Here $N_f(\vec{x})$ is the number
density of fermion $f$ at the position $\vec{x}$ along the neutrino
trajectory.  In Eq.~\eqref{eq:epx-nsi} we have limited the sum to the
charged fermions present in ordinary matter, $f=e,u,d$.  Taking into
account that $N_u(\vec{x}) = 2N_p(\vec{x}) + N_n(\vec{x})$ and
$N_d(\vec{x}) = N_p(\vec{x}) + 2N_n(\vec{x})$, and also that matter
neutrality implies $N_p(\vec{x}) = N_e(\vec{x})$,
Eq.~\eqref{eq:epx-nsi} becomes:
\begin{equation}
  \label{eq:epx-nuc}
  \Epx_{\alpha\alpha}(\vec{x}) =
  \big( \Eps_{\alpha\alpha}^e + \Eps_{\alpha\alpha}^p \big)
  + Y_n(\vec{x}) \Eps_{\alpha\alpha}^n
  \quad\text{with}\quad
  Y_n(\vec{x}) \equiv \frac{N_n(\vec{x})}{N_e(\vec{x})}
\end{equation}
where
\begin{equation}
  \label{eq:eps-nucleon}
  \Eps_{\alpha\alpha}^p \equiv 2\Eps_{\alpha\alpha}^u + \Eps_{\alpha\alpha}^d
  = a_p\, b_\alpha \, \Eps^0 \,,
  \qquad
  \Eps_{\alpha\alpha}^n \equiv 2\Eps_{\alpha\alpha}^d + \Eps_{\alpha\alpha}^u
  = a_n \,b_\alpha\, \Eps^0 \,.
\end{equation}
and we have introduced the proton and neutron $Z'$ couplings
\begin{equation}
  a_p \equiv 2 a_u + a_d \,,
  \qquad
  a_n \equiv 2 a_d +a_u \,.
\end{equation}
As discussed in Ref.~\cite{GonzalezGarcia:2011my}, in the Earth the
neutron/proton ratio $Y_n(\vec{x})$ which characterize the matter
chemical composition can be taken to be constant to very good
approximation.  The PREM model~\cite{Dziewonski:1981xy} fixes $Y_n =
1.012$ in the Mantle and $Y_n = 1.137$ in the Core, with an average
value $Y_n^\oplus = 1.051$ all over the Earth. Setting therefore
$Y_n(\vec{x}) \equiv Y_n^\oplus$ in Eqs.~\eqref{eq:epx-nsi}
and~\eqref{eq:epx-nuc} we get $\Epx_{\alpha\alpha}(\vec{x}) \equiv
\Eps_{\alpha\alpha}^\oplus$ with:
\begin{equation}
  \begin{split}
    \Eps_{\alpha\alpha}^\oplus
    &= \Eps_{\alpha\alpha}^e + \big( 2 + Y_n^\oplus \big) \Eps_{\alpha\alpha}^u
    + \big( 1 + 2Y_n^\oplus \big) \Eps_{\alpha\alpha}^d
    = \big( \Eps_{\alpha\alpha}^e + \Eps_{\alpha\alpha}^p \big)
    + Y_n^\oplus \Eps_{\alpha\alpha}^n
    \\
    &= \big[(a_e + a_p) + Y_n^\oplus a_n \big]\, b_\alpha\, \Eps^0 \,.
\end{split}
\end{equation}

For what concerns the study of propagation of solar and KamLAND
neutrinos one can work in the one mass dominance approximation,
$\Dmq_{31} \to \infty$ (which effectively means that $G_F N_e(\vec{x})
\Epx_{\alpha\alpha}(\vec{x}) \ll \Dmq_{31} / E_\nu$). In this
approximation the survival probability $P_{ee}$ can be written
as~\cite{Kuo:1986sk, Guzzo:2000kx}
\begin{equation}
  \label{eq:peesun}
  P_{ee} = c_{13}^4 P_\text{eff} + s_{13}^4
\end{equation}
The probability $P_\text{eff}$ can be calculated in an effective
$2\times 2$ model described by the Hamiltonian $H_\text{eff} =
H_\text{vac}^\text{eff} + H_\text{mat,SM}^\text{eff} +
H_\text{mat,$Z'$}^\text{eff}$, with:
\begin{align}
  \label{eq:HvacSol}
  H_\text{vac}^\text{eff}
  &= \frac{\Dmq_{21}}{4 E_\nu}
  \begin{pmatrix}
    -\cos2\theta_{12} \, \hphantom{e^{-i\delta_\text{CP}}}
    & ~\sin2\theta_{12} \, e^{i\delta_\text{CP}}
    \\
    \hphantom{-}\sin2\theta_{12} \, e^{-i\delta_\text{CP}}
    & ~\cos2\theta_{12} \, \hphantom{e^{i\delta_\text{CP}}}
  \end{pmatrix},
  \\
  \label{eq:HmatSolSM}
  H_\text{mat,SM}^\text{eff}
  &= \sqrt{2} G_F N_e(\vec{x})
  \begin{pmatrix}
    c_{13}^2 & 0 \\
    0 & 0
  \end{pmatrix}
\end{align}
and
\begin{equation}
  \label{eq:HvacZp}
  H_\text{mat,Z'}^\text{eff}
  = \sqrt{2} G_F N_e(\vec{x})\, \Eps^0 \,
  \big[ a_e + a_p + Y_n(\vec{x}) a_n \big]
  \begin{pmatrix}
    -b_D & b_N \\
    \hphantom{+} b_N & b_D
  \end{pmatrix}
\end{equation}
where
\begin{align}
  \label{eq:b_D}
  \begin{split}
   b_D &=
    -\frac{c_{13}^2}{2} \left(b_e - b_\mu\right)
    + \frac{s_{23}^2 - s_{13}^2 c_{23}^2}{2}
    \left(b_\tau - b_\mu\right) \,,
  \end{split}
  \\[2mm]
  \label{eq:b_N}
  b_N &= s_{13} c_{23} s_{23} \left(b_\tau - b_\mu\right) \,.
\end{align}
Following Ref.~\cite{Esteban:2018ppq} we can rewrite the $Z'$
contribution as:
\begin{align}
\label{eq:HmatZp2}
  H_\text{mat,Z'}^\text{eff}
  &= \sqrt{2} G_F N_e(\vec{x})
   \big[\cos\eta + Y_n(\vec{x}) \sin\eta \big]
    \begin{pmatrix}
      -\Eps_D^{\eta\hphantom{*}} & \Eps_N^\eta \\
      \hphantom{+} \Eps_N^{\eta} & \Eps_D^\eta
    \end{pmatrix},
\end{align}
where the angle $\eta$ parametrizes the ratio of the charges of the
matter particles as:
\begin{equation}
  \cos\eta = \frac{a_e + a_p}{\sqrt{(a_e + a_p)^2+a_n^2}} \,,
  \qquad
  \sin\eta = \frac{a_n}{\sqrt{(a_e + a_p)^2+a_n^2}} \,,
\end{equation}
and
\begin{equation}
  \Eps_{D,N}^\eta = \sqrt{(a_e + a_p)^2 + a_n^2}\, b_{D,N}\, \Eps^0 \,.
\end{equation}

The neutrino oscillation phenomenology in this regime reduces to a
special subclass of the general NSI interactions analyzed in
Ref.~\cite{Esteban:2018ppq}.\footnote{To be precise, the data analysis
performed in Ref.~\recite{Esteban:2018ppq} was restricted to NSI with
quarks, ie $a_e=0$. The formalism for matter effects can be trivially
extended to NSI coupled to electrons as shown above.  However, NSI
coupled to electrons would affect not only neutrino propagation in
matter as described, but also the neutrino-electron (ES) scattering
cross-section in experiments such as SK, SNO and Borexino. In order to
keep the analysis manageable, in Ref.~\recite{Esteban:2018ppq}, and in
what follows, the NSI corrections to the ES scattering cross section
in SK, SNO, and Borexino are neglected. In the absence of
cancellations between propagation and interaction effects this renders
the results of the oscillation analysis conservative.}  In particular,
as a consequence of the CPT symmetry (see also
Refs.~\cite{GonzalezGarcia:2011my, Gonzalez-Garcia:2013usa,
  Coloma:2016gei} for a discussion in the context of NSI) the neutrino
evolution is invariant if the relevant Hamiltonian is transformed as
$H \to -H^*$. In vacuum this transformation can be realized by
changing the oscillation parameters as
\begin{equation}
  \label{eq:osc-deg}
  \begin{aligned}
    \Dmq_{31} &\to -\Dmq_{31} + \Dmq_{21} = -\Dmq_{32} \,, \\
    \theta_{12} &\to \pi - \theta_{12} \,,\\
    \delta_\text{CP} &\to \pi - \delta_\text{CP} \,,
  \end{aligned}
\end{equation}
where $\delta_\text{CP}$ is the leptonic Dirac CP phase, and we are
using here the parameterization conventions from
Refs.~\cite{Esteban:2018ppq, Coloma:2016gei}.  The symmetry is broken
by the standard matter effect, which allows a determination of the
octant of $\theta_{12}$ and (in principle) of the sign of
$\Dmq_{31}$. However, in the presence of the $Z'$-induced NSI, the
symmetry can be restored if in addition to the transformation
Eq.~\eqref{eq:osc-deg}, the $\Epx_{\alpha\alpha}(\vec{x})$ terms can
be transformed as~\cite{Gonzalez-Garcia:2013usa, Bakhti:2014pva,
  Coloma:2016gei}
\begin{equation}
  \label{eq:NSI-deg}
  \begin{aligned}
    \big[ \Epx_{ee}(\vec{x}) - \Epx_{\mu\mu}(\vec{x}) \big]
    &\to - \big[ \Epx_{ee}(\vec{x}) - \Epx_{\mu\mu}(\vec{x}) \big] - 2  \,,
    \\
    \big[ \Epx_{\tau\tau}(\vec{x}) - \Epx_{\mu\mu}(\vec{x}) \big]
    &\to -\big[ \Epx_{\tau\tau}(\vec{x}) - \Epx_{\mu\mu}(\vec{x}) \big] \,.
  \end{aligned}
\end{equation}
Eq.~\eqref{eq:osc-deg} shows that this degeneracy implies a change in
the octant of $\theta_{12}$ (as manifest in the LMA-D fit to solar
neutrino data~\cite{Miranda:2004nb}) as well as a change in the
neutrino mass ordering, \textit{i.e.}, the sign of $\Dmq_{31}$. For
that reason it has been called ``generalized mass ordering
degeneracy'' in Ref.~\cite{Coloma:2016gei}.  Because of the position
dependence of the NSI hamiltonian described by
$\Epx_{\alpha\alpha}(\vec{x})$ this degeneracy is only approximate,
mostly due to the non-trivial neutron/proton ratio along the neutrino
path inside the Sun. In what follows when marginalizing over
$\theta_{12}$ we consider two distinct parts of the parameter space:
one with $\theta_{12} < 45^\circ$, which we denote as $\text{LIGHT}$,
and one with $\theta_{12} > 45^\circ$, which we denote by
$\text{DARK}$.

Apart from the appearance of this degenerate solution, another feature
to consider in the global analysis of oscillation data in presence of
NSI is the possibility to further improve the quality of the fit with
respect to that of \emph{standard} $3\nu$ oscillations in the
$\text{LIGHT}$ sector. Till recently this was indeed the case because
for the last decade the value of $\Dmq_{21}$ preferred by KamLAND was
somewhat higher than the one from solar experiments.  This tension
appeared due to a combination of two effects: the fact that the
\Nuc{8}{B} measurements performed by SNO, SK and Borexino showed no
evidence of the low energy spectrum turn-up expected in the standard
LMA-MSW~\cite{Wolfenstein:1977ue, Mikheev:1986gs} solution for the
value of $\Dmq_{21}$ favored by KamLAND, and the observation of a
non-vanishing day-night asymmetry in SK, whose size was larger than
the one predicted for the $\Dmq_{21}$ value indicated by KamLAND.
With the data included in the analysis in Ref.~\cite{Esteban:2018azc,
  Esteban:2018ppq} this resulted into a tension of $\Delta\chi^2\sim
7.4$ for the standard $3\nu$ oscillations.  Such tension could be
alleviated in presence of a non-standard matter potential, thus
leading to a possible decrease in the minimum $\chi^2$.  However, with
the latest 2970-days SK4 results presented at the Neutrino2020
conference~\cite{SK:nu2020} in the form of total energy spectrum and
updated day-night asymmetry, the tension between the best fit
$\Dmq_{21}$ of KamLAND and that of the solar results has
decreased. Currently they are compatible within $1.1\sigma$ in the
latest global analysis~\cite{Esteban:2020cvm}.

\subsection{The finite $M_{Z'}$ case: the long-range interaction regime}
\label{sec:LR}

If $M_{Z'}$ is very light the four-fermion contact interaction
approximation in Eq.~\eqref{eq:NSILagrangian} does not hold and the
potential encountered by the neutrino in its trajectory depends on the
integral of the source density within a radius $\sim 1/M_{Z'}$ around
it. However, following Ref.~\cite{GonzalezGarcia:2006vp} the
generalized matter potential can still be written as
Eq.~\eqref{eq:Hmat} provided that Eq.~\eqref{eq:epx-nsi} is modified
as:
\begin{equation}
  \label{eq:epx-lri}
  \Epx_{\alpha\alpha}(\vec{x}) = \sum_{f=e,u,d}
  \frac{\hat{N}_f(\vec{x},M_{Z'})}{N_e(\vec{x})} \Eps_{\alpha\alpha}^f
\end{equation}
where
\begin{equation}
  \hat{N}_f(\vec{x},M_{Z'}) \equiv \frac{M_{Z'}^2}{4\pi}
  \int N_f(\vec\rho)\,
  \frac{e^{-M_{Z'} |\vec\rho - \vec{x}|}}{|\vec\rho - \vec{x}|} \,
  d^3\vec\rho \,.
\end{equation}
Taking into account that ordinary matter is neutral and only contains
$f=e,u,d$, we can rewrite Eq.~\eqref{eq:epx-lri} in a way that
generalizes Eq.~\eqref{eq:epx-nuc}:
\begin{multline}
  \Epx_{\alpha\alpha}(\vec{x}) =
  F_e(\vec{x},M_{Z'}) \big( \Eps_{\alpha\alpha}^e + \Eps_{\alpha\alpha}^p \big)
  + F_n(\vec{x},M_{Z'}) Y_n(\vec{x}) \Eps_{\alpha\alpha}^n
  \\
  \text{with}\quad
  F_i(\vec{x},M_{Z'}) \equiv \frac{\hat{N}_i(\vec{x},M_{Z'})}{N_i(\vec{x})}
  \quad\text{and}\quad
  i \in \lbrace e, n \rbrace \,.
\end{multline}
For what concerns neutrinos traveling inside the Sun, the propagation
effects induced by the new interactions are completely dominated by
the solar matter distribution. Denoting by $\vec{x}_\odot$ the center
of the Sun and accounting for the spherical symmetry of the matter
potential we can write:
\begin{multline}
  F_i(\vec{x},M_{Z'}) \simeq F_i^\odot(|\vec{x}-\vec{x}_\odot|, M_{Z'})
  \\
  \text{with}\quad
  F_i^\odot(r,M_{Z'})
  = \frac{1}{N_i^\odot(r)} \cdot \frac{M_{Z'}}{2\,r}
  \int_{0}^{R_\odot} \rho\, N_i^\odot(\rho)
  \big[e^{-M_{Z'}|\rho-r|} - e^{-M_{Z'}(\rho+r)} \big] \, d\rho \,.
\end{multline}
A similar formula can be derived for neutrinos traveling inside the
Earth, but in this case the effective potential has an extra term
induced by the Sun matter density. Concretely, denoting by
$\vec{x}_\oplus$ the center of the Earth and by $X_\ominus =
|\vec{x}_\odot - \vec{x}_\oplus|$ the Sun-Earth distance, we have:
\begin{multline}
  F_i(\vec{x},M_{Z'}) \simeq F_i^\oplus(|\vec{x}-\vec{x}_\oplus|, M_{Z'})
  \\
  \text{with}\quad
  F_i^\oplus(r,M_{Z'})
  = \frac{1}{N_i^\oplus(r)} M_{Z'} \Bigg\lbrace
  \frac{1}{2\,r}
  \int_{0}^{R_\oplus} \rho\, N_i^\oplus(\rho)
  \big[e^{-M_{Z'}|\rho-r|} - e^{-M_{Z'}(\rho+r)} \big] \, d\rho
  \\
  + \frac{e^{-M_{Z'} X_\ominus}}{X_\ominus}
  \int_{0}^{R_\odot} \rho\, N_i^\odot(\rho)
  \sinh(\rho\, M_{Z'}) \, d\rho \Bigg\rbrace \,.
\end{multline}
The solar-induced contribution becomes non-negligible when the range
of the interactions, $1\big/M_{Z'}$, is comparable or larger than the
Sun-Earth distance $X_\ominus$.

The factors $F_i^\odot(r,M_{Z'})$ and $F_i^\oplus(r,M_{Z'})$ represent
the modification of the matter potential due to the finite range of
the interaction mediated by the $Z'$ with respect to that obtained in
the contact interaction limit. Such limit is recovered when the range
of the new interactions become shorter than the typical size of the
matter distribution, \textit{i.e.}, $R_{\odot(\oplus)}$. Hence:
\begin{equation}
  F_i^{\odot(\oplus)} (r, M_{Z'}) \to 1
  \quad\text{for}\quad
  M_{Z'}\gg 1/R_{\odot(\oplus)} \,.
\end{equation}
For solar neutrinos further simplification follows if one takes into
account that for adiabatic transitions the dominant matter effects is
generated by the potential at the neutrino production point which is
is close to the Sun center. So to a very good approximation one can
scale the contact interaction potential with a position independent
factor $F_i^\odot(0,M_{Z'})$.
For the Earth matter potential the position dependence of the factor
$F_i^\oplus(r,M_{Z'})$ is very weak in the current experiments, so one
can also scale the contact interaction potential with an approximate
$F_i^\oplus(\bar{r},M_{Z'})$ evaluated at a fix $\bar{r}$ which we
take to be also $\bar{r} = 0$.

\begin{figure}\centering
  \includegraphics[width=0.49\textwidth]{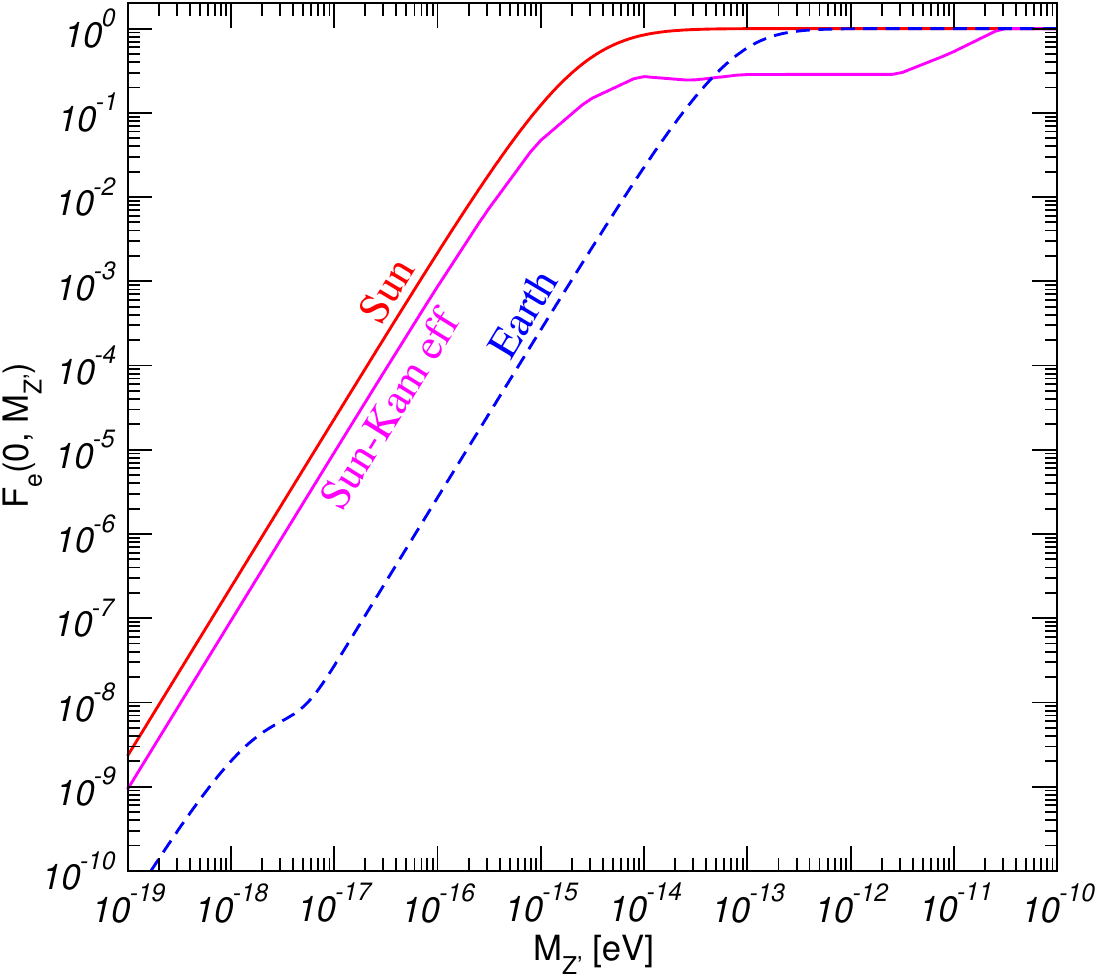}
  \caption{Effective suppression factor of the matter potential due to
    the long range of the $U(1)'$ interactions as a function of
    $M_{Z'}$. The red (blue) curve corresponds to the potential in the
    Sun (Earth). The purple line corresponds to the effective combined
    suppression factor in the analysis of Solar+KamLAND data (see text
    for details).}
  \label{fig:scaling}
\end{figure}

In Fig.~\ref{fig:scaling} we plot these scale factors
$F_e^{\odot(\oplus)}(0,M_{Z'})$. As seen in the figure for
$M_{Z'}\lesssim 10^{-13}$~eV the matter potential in the Earth is more
suppressed with respect to that in the Sun. In principle, this opens
the possibility of configurations for which the $U(1)'$-induced matter
potential in the Sun is large enough without conflicting with bounds
imposed by atmospheric and long-baseline experiments.  This also
implies that in the combined analysis of solar and KamLAND data, for a
given value of $M_{Z'}$, the effective matter potential for solar
neutrinos will be suppressed by a different factor than that for
KamLAND antineutrinos. To illustrate the overall $M_{Z'}$ dependence
of the effect we show in Fig.~\ref{fig:scaling} the effective
suppression factor in the combined solar+KamLAND analysis calculated
by scaling the results obtained for a specific model (concretely, for
a $Z'$ coupled to $L_e-L_\mu$, but the results are similar for models
with other couplings) for each $M_{Z'}$ to those obtained in the large
$M_{Z'}$ regime.  As seen in the figure for $M_{Z'}\gtrsim
10^{-10}$~eV both the effective potential in the Solar+KamLAND
analysis and the Earth matter potential relevant for atmospheric and
LBL neutrinos are well within the contact interaction
regime. Conversely for $M_{Z'}\lesssim 10^{-13}$~eV all the matter
potentials in the analysis show deviations from the contact
interaction regime.

Concerning the phenomenology of neutrino oscillations in the presence
of the modified matter potential in this long-range interaction
regime, the main difference with the NSI contact interaction case is
the impossibility of realizing the ``generalized mass ordering
degeneracy'' in Eqs.~\eqref{eq:osc-deg} and~\eqref{eq:NSI-deg} because
of the very different $\vec{x}$ dependence of the SM matter potential
and the one generated by the $Z'$. In other words, one cannot ``flip
the sign of the matter hamiltonian'' by adding to the standard
$N_e(x)$ something which has a completely different $\vec{x}$
profile. Hence, there is no LMA-D solution for these models.
On the other hand, it is still possible, at least in principle, that a
long-range potential leads to an improvement on the fit to solar and
KamLAND data with respect to the pure LMA solution.

\section{Results of the global oscillation analysis}
\label{sec:results}

We have performed a global fit to neutrino oscillation data in the
framework of $3\nu$ massive neutrinos with new neutrino-matter
interactions generated by $U(1)'$ models and characterized by the
Lagrangian in Eq.~\eqref{eq:zpcurrent}.  For the detailed description
of methodology and data included we refer to the comprehensive global
fit in Ref.~\cite{Esteban:2018ppq} performed in the framework of
three-flavor oscillations plus NSI. In addition in the present
analysis we account for the latest LBL data samples included in
NuFIT-5.0~\cite{Esteban:2020cvm} which includes the previously cited
solar neutrinos 2970-days SK4 results~\cite{SK:nu2020}, the updated
medium baseline reactor samples from RENO~\cite{RENO:nu2020} and
Double Chooz~\cite{DoubleC:nu2020}, and the latest long-baseline
samples from T2K~\cite{T2K:nu2020} and NO$\nu$A~\cite{NOvA:nu2020}.
Notice that in order to keep the fit manageable we proceed as in in
Ref.~\cite{Esteban:2018ppq} and restrict ourselves to the
CP-conserving case and set $\delta_\text{CP} \in \{ 0, \pi \}$.
Consequently the T2K and NO$\nu$A appearance data (which exhibit
substantial dependence on the leptonic CP phase) are not included in
the fit.  With these data we construct a global $\chi^2$ function:
\begin{equation}
  \chi^2_\text{OSC+Z'} (g',M_{Z'}|\vec \omega)
\end{equation}
for each $Z'$ model. In general each model belongs to a family
characterized by a set of $U(1)'$ charges; for each family
$\chi^2_\text{OSC+Z'}$ depends on the two variables parametrizing the
new interaction, $g'$ and $M_{Z'}$, plus the six oscillation
parameters $\vec\omega\equiv (\Dmq_{21}, \Dmq_{31}, \theta_{12},
\theta_{13}, \theta_{23}, \delta_\text{CP})$.

Following the discussion in Sec.~\ref{sec:forma}, we have performed
the analysis in two physically distinctive domains of $M_{Z'}$:
\begin{itemize}
\item{NSI Domain (DOM=NSI)}: In this case, the range of the induced
  non-standard interactions in both the Sun and the Earth matter is
  short enough for the four-fermion effective description to hold.  As
  seen in Fig.~\ref{fig:scaling} this happens for $M_{Z'} \gtrsim
  10^{-10}$~eV. In this regime a possible conflict with the
  cosmological bound on $\Delta N_\text{eff}$ may appear because of
  the contribution of either the $Z'$ itself (if lighter than all
  active neutrinos) or by the extra contribution to the neutrino
  density produced by the $Z'$ decay (if heavier than some $\nu$ mass
  eigenstate).  These bounds can be evaded in two distinct ranges of
  the $U(1)'$ interactions:
  \begin{itemize}
  \item $M_{Z'} \gtrsim 5$~MeV for which the contribution to the
    neutrino energy density due to the decay of the $Z'$ into
    neutrinos is sufficiently suppressed by the Boltzmann factor $\sim
    \exp(-M_{Z'}/T)$~\cite{Kamada:2015era};

  \item $M_{Z'} \lesssim \mathcal{O}(\text{eV})$ but with very weak
    coupling $g'<10^{-10}$ for which the $Z'$ is produced through
    freeze-in.  In this regime, even if $Z'$ could decay to the
    lightest neutrinos this would happen after neutrino decoupling
    making the contribution to $\Delta N_\text{eff}$ negligible or at
    most within the present allowed range~\cite{Escudero:2019gzq}.
  \end{itemize}

\item{Long-Range Domain (DOM=LRI)}: If $M_{Z'}\lesssim 10^{-13}$~eV
  interactions in the Earth and Sun matter are long range, and for a
  given value of $g'$ and $M_{Z'}$ the effects in the Earth are
  suppressed with respect to those in the Sun. As mentioned above, for
  $g'<10^{-10}$ the contribution to $\Delta N_\text{eff}$ is
  negligible.
\end{itemize}
Correspondingly we define
\begin{equation}
  \chi^2_\text{OSC+Z',DOM} (g',M_{Z'}|\vec\omega)\equiv
  \chi^2_\text{OSC+Z'} (g',M_{Z'}\in \text{DOM}|\vec\omega)
\end{equation}
In both domains we compare the results of the fit including the new
$U(1)'$ interaction with those obtained in the ``standard''
$3\nu$-mixing scenario, which we will denote as ``OSC'', and for which
the present global fit yields
\begin{equation}
  \chi^2_\text{OSC,min} = 718.5 \,.
\end{equation}

We will classify the models according to the quality of the fit in the
presence of the $U(1)'$ interactions compared with that of OSC by
defining
\begin{equation}
  \Delta\chi^2_\text{LIGHT,DOM}(g',M_{Z'})
  \equiv \left. \chi^2_\text{OSC+Z',DOM}(g',M_{Z'}|\vec\omega)
  \right|_\text{marg,LIGHT} - \chi^2_\text{OSC,min} \,,
  \label{eq:dchi2ldef}
\end{equation}
where by $\left.\right|_\text{marg,LIGHT}$ we imply that the
minimization over the oscillation parameters is done in the
$\text{LIGHT}$ sector of parameter space.
In addition the presence of a viable LMA-D solution can be quantified
in terms of
\begin{equation}
  \Delta\chi^2_\text{DARK,DOM}(g',M_{Z'})
  \equiv \left.\chi^2_\text{OSC+Z',DOM}(g',M_{Z'}|\vec\omega)
  \right|_\text{marg, DARK} - \chi^2_\text{OSC,min} \,,
\end{equation}
where by $\left.\right|_\text{marg,DARK}$ we imply that the
minimization over the oscillation parameters is done in the
$\text{DARK}$ sector.

We have surveyed the model space by performing the global oscillation
analysis for a grid of $U(1)'$ interactions characterized by the six
couplings $a_{u,d} \in \{-1, 0, 1\}$ and $a_e,\, b_{e,\mu,\tau} \in
\{-3, -2, -1, 0, 1, 2, 3\}$.  In this way our survey covers a total of
$\sim 10000$ different sets of $U(1)'$ charges which can produce
effects in matter propagation in neutrino oscillation experiments.

\subsection{Bounds}
\label{sec:bounds}

We first search for models for which in the LIGHT sector the new
interactions lead to a significantly better fit of the oscillation
data compared to pure oscillations for some value of $g'$ and
$M_{Z'}$. We find that in the NSI (LRI) domain 88\% (90\%) of the
surveyed sets of charges lead to a decrease in the $\chi^2$ of the
global analysis when compared to the standard oscillation case. The
percentages grow to 100\% when restricting to the subclass of anomaly
free vector $Z'$ models with gauging of SM global symmetries with SM
plus right-handed neutrinos matter content of Eq.~\eqref{eq:anomod}.
However the improvement in the quality of the fit is never
statistically significant.  As an illustration we show in the second
column of tables~\ref{tab:boundsLRI} and~\ref{tab:boundsNSI} the
minimum values of $\Delta\chi^2_\text{LIGHT,DOM}$ for $U(1)'$
interactions characterized by the specific set of charges in
table~\ref{tab:models}. Comparing the two tables we notice that for
some of the cases the fit can be slightly better in the LRI domain
than in the NSI domain but still below the $2\sigma$ level.

\begin{table}\centering
  \catcode`!=\active\def!{\hphantom{-}}
  \catcode`?=\active\def?{\hphantom{0}}
  \begin{tabular}{@{\rule[-2.1mm]{0pt}{6mm}}|c|c|c|}
    \hline
    Model
    & $(\Delta\chi_\text{LIGHT,LRI}^2)_\text{min}$ & $g'\leq$ bound
    \\ \hline
    $B - 3L_e$
    & $-1.4$
    &$6.6\times 10^{-27}$
    \\
    $B - 3L_\mu$
    & $-1.1$
    & $7.0\times 10^{-27}$
    \\
    $B - 3L_\tau$
    & $-1.8$
    & $7.3\times 10^{-27}$
    \\
    $B - \frac{3}{2}(L_\mu + L_\tau)$
    & $-1.2$
    &$7.2\times 10^{-27}$
    \\
    $L_e - L_\mu$
    & $-1.3$
    &$9.7\times 10^{-27}$
    \\
    $L_e - L_\tau$
    & $-1.7$
    & $1.0\times 10^{-26}$
    \\
    $L_e - \frac{1}{2}(L_\mu + L_\tau)$
    & $-1.4$
    & $9.8\times 10^{-27}$
    \\
    $B_y + L_\mu + L_\tau$ Ref.~\recite{Farzan:2015doa}
    & $!0\hphantom{.0}$
    & $4.9\times 10^{-27}$
    \\
    $L_e + 2L_\mu + 2L_\tau$
    & $!0\hphantom{.0}$
    & $6.0\times 10^{-27}$
    \\
    \hline
  \end{tabular}
  \caption{Results for the models with charges in
    table~\ref{tab:models} in the LRI regime.  For the model in
    Ref.~\recite{Farzan:2015doa} we have defined $B_y \equiv B_1 - y
    B_2 - (3-y)B_3$.  The second column gives minimum $\Delta\chi^2$
    defined w.r.t.\ the $3\nu$ oscillation (see
    Eq.~\eqref{eq:dchi2ldef}). The last column gives the the upper
    bound for the coupling of asymptotically for ultra light
    mediators, $M_{Z'}\lesssim 10^{-15}$~eV}
  \label{tab:boundsLRI}
\end{table}

Quantitatively we find that in the NSI regime none of the surveyed
models yields an improvement beyond -1.9 units of $\chi^2$. This is
the case, for example, of a $Z'$ coupled to $B - 3L_e + 2L_\mu +
3L_\tau$.  Generically models in the LRI domain can provide a better
fit with a reduction of up to 3.5 units of $\chi^2$. For example a
model with charge $L_e + 2L_\mu - 3L_\tau$ and a $Z'$ with $M_{Z'}\sim
5\times 10^{-15}$~eV provides a better fit than standard oscillations
by $(\Delta\chi^2_\text{LIGHT,LRI})_\text{min}=-3.2$.

\begin{table}\centering
  \catcode`!=\active\def!{\hphantom{-}}
  \catcode`?=\active\def?{\hphantom{0}}
  \begin{tabular}{@{\rule[-2.1mm]{0pt}{6mm}}|c|c|c|}
    \hline
    Model
    & $(\Delta\chi_\text{LIGHT,NSI}^2)_\text{min}$ & $g'\leq$ bound
    $\bigg( \dfrac{M_{Z'}}{\text{100~MeV}} \bigg)$
    \\ \hline
    $B - 3L_e$
    & $-1.4$
    & $2.0\times 10^{-4}$
    \\
    $B - 3L_\mu$
    & $!0\hphantom{.0}$
    & $4.6\times 10^{-5}$
    \\
    $B - 3L_\tau$
    & $-0.6$
    & $4.7\times 10^{-5}$
    \\
    $B - \frac{3}{2}(L_\mu + L_\tau)$
    & $-1.1$
    & $2.2\times 10^{-4}$
    \\
    $L_e - L_\mu$
    & $-1.3$
    & $1.2\times 10^{-4}$
    \\
    $L_e - L_\tau$
    & $-1.0$
    & $1.2\times 10^{-4}$
    \\
    $L_e - \frac{1}{2}(L_\mu + L_\tau)$
    & $-1.3$
    & $3.0\times 10^{-4}$
    \\
    $B_y + L_\mu + L_\tau$ Ref.~\recite{Farzan:2015doa}
    & $!0\hphantom{.0}$
    & $1.5\times 10^{-4}$
    \\
    $L_e + 2L_\mu + 2L_\tau$
    & $-0.1$
    & $1.8\times 10^{-4}$
    \\
    \hline
  \end{tabular}
  \caption{Results for the models with charges in
    table~\ref{tab:models} in the NSI regime. For the model in
    Ref.~\recite{Farzan:2015doa} we have defined $B_y \equiv B_1 - y
    B_2 - (3-y)B_3$. The second column gives minimum $\Delta\chi^2$
    defined w.r.t.\ the $3\nu$ oscillation (see
    Eq.~\eqref{eq:dchi2ldef}). The last column gives the coefficient
    of the bound on the coupling over the mediator mass in units of
    100 MeV.}
  \label{tab:boundsNSI}
\end{table}

But in summary, our analysis shows that none of the set of charges
surveyed in both NSI or LRI domains improved over standard
oscillations at the $2\sigma$ level, this is
$\min_{g',M_{Z'}}(\Delta\chi^2_\text{LIGHT,DOM})$ was always larger
than $-4$.\footnote{We notice that before the new results from
Super-Kamiokande~\recite{SK:nu2020} there were models which could
improve the mismatch between the best fit $\Dmq_{21}$ in Solar and
KamLAND.  For such models, one could find values of $g'$ and $M_{Z'}$
for which the fit was better than standard oscillations by more than 4
units of $\chi^2$.}  Consequently for all models surveyed one can
conclude that the analysis of neutrino oscillation experiments show no
significant evidence of $U(1)'$ interactions.  Consequently one can
exclude models at a certain confidence level --~which we have chosen
to be 95.45\%~-- by verifying that their global fit is worse than in
OSC by the corresponding units of $\chi^2$ (4 units for 95.45\%~CL),
this is:
\begin{equation}
  \label{eq:dchibound}
  \Delta\chi^2_\text{LIGHT,DOM}(g',M_{Z'}) > 4.
\end{equation}
Let us stress that with the above condition we are not ``deriving
two-dimensional excluded regions in the parameter space'', but we are
instead determining the values of $g'$ and $M_{Z'}$ for which the
$U(1)'$ model characterized by such interaction strength and
interaction length, gives a fit which is worse than standard
oscillations by at least 4 units of $\chi^2$. As in the LIGHT sector
the standard model is recovered for either $g'\to 0$ or $M_{Z'}\to
\infty$, the above condition yields also the $2\sigma$ excluded
one-dimensional upper range of interaction coupling $g'$ for each
value of the interaction length (or, correspondingly, the $2\sigma$
excluded one-dimensional lower range of $M_{Z'}$ for each value of the
interaction coupling), for all models characterized by a given set of
charges.

\begin{pagefigure}\centering
  \includegraphics[height=0.79\textwidth,angle=90]{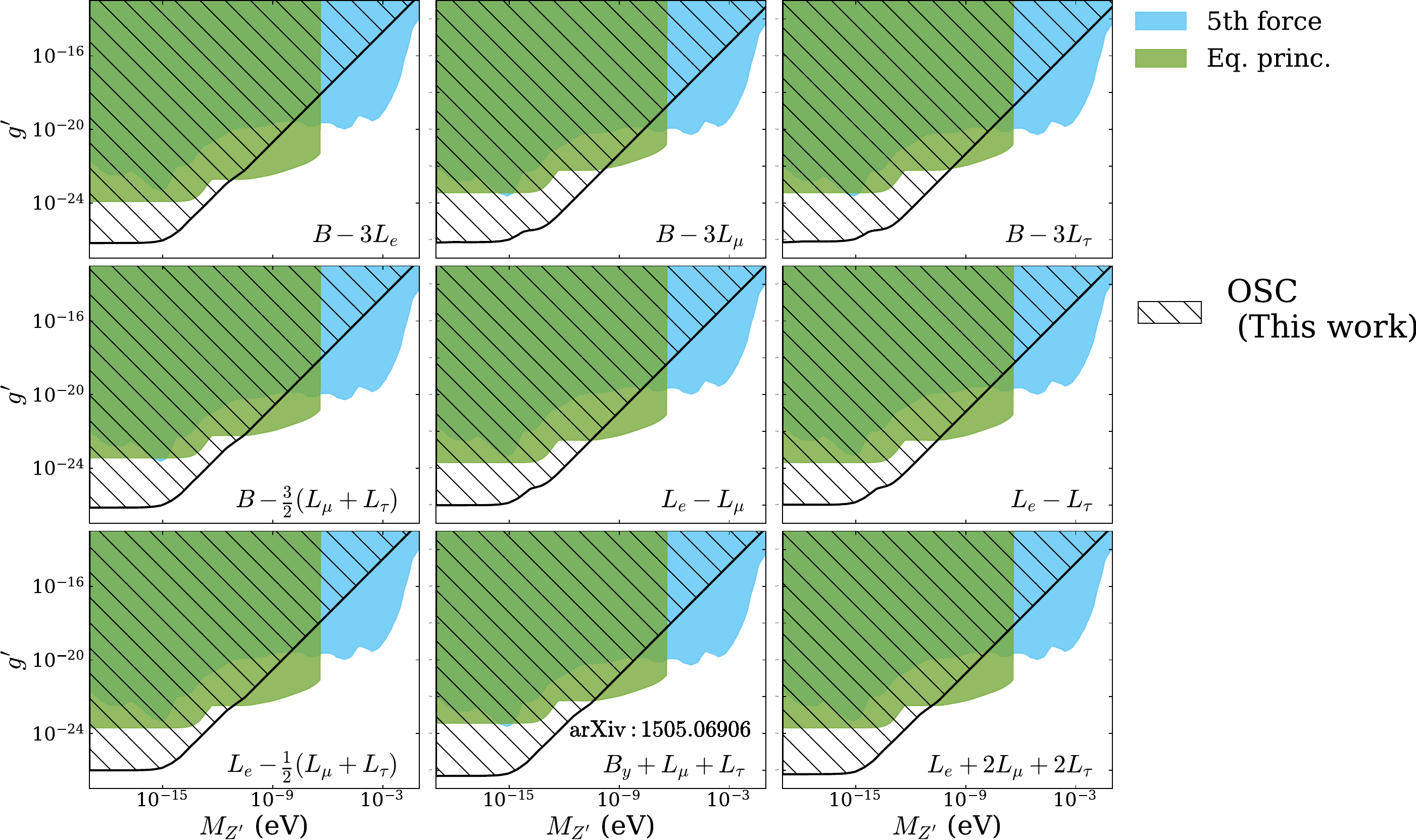}
  \caption{Values of $g'$ and $M_{Z'}\leq \mathcal{O}$(eV) for which a
    $U(1)'$ model coupled to the charges labeled in each panel gives a
    worse fit than standard oscillation by $2\sigma$,
    Eq.~\eqref{eq:dchibound} (hatched region).  We also show the
    bounds on these models imposed by gravitational fifth force
    searches~\recite{Salumbides:2013dua, Adelberger:2009zz} and by
    equivalence principle tests~\recite{Schlamminger:2007ht}.  In the
    window corresponding to the model in Ref.~\recite{Farzan:2015doa}
    we have defined $B_y \equiv B_1 - y B_2 - (3-y)B_3$.  See text for
    details.}
  \label{fig:lowmass}
\end{pagefigure}

The corresponding excluded ranges for the $Z'$ coupling and mass for
the models with couplings listed in table~\ref{tab:models} are shown
in Figs.~\ref{fig:lowmass} and~\ref{fig:highmass} for $M_{Z'}$ below
1~eV and in the $\mathcal{O}(\text{MeV--GeV})$ range,
respectively\footnote{Tables with the numerical values of the bounds
can be provided upon request to the authors.} -- \textit{i.e.}, below
and above the window strongly disfavored by the cosmological bound on
$\Delta N_\text{eff}$.
In particular in Fig.~\ref{fig:lowmass} we observe the slope change of
the oscillation exclusion ranges for masses $M_{Z'}\sim
10^{-15}$--$10^{-13}$~GeV, for which the interaction length is longer
than the Earth and Sun radius and the matter potential in the Earth
and in the Sun becomes saturated (see Fig.~\ref{fig:scaling}).
Quantitatively, for the models with charges in table~\ref{tab:models}
we find that in the LRI domain the analysis of oscillation data yields
the upper bound for the coupling of asymptotically ultra light
mediators, $M_{Z'}\lesssim 10^{-15}$~eV which we list in
table~\ref{tab:boundsLRI}.
For the sake of comparison, we also show in Fig.~\ref{fig:lowmass} the
bounds on these models imposed by gravitational fifth force searches,
and by equivalence principle tests.  Those are the strongest model
independent constraints in the shown range of $Z'$ mass and coupling
derived at comparable confidence level under the minimal assumptions
in Eq.~\eqref{eq:zpcurrent}.  For some of the models shown in the
figure, additional bounds on this range of masses can arise from
cosmological and astrophysical observations, including constraints on
invisible neutrino decay $\nu_a\to \nu_b Z'$ from Cosmic Microwave
Background (CMB) data~\cite{Escudero:2019gfk}, bounds from Black-Hole
superradiance~\cite{Baryakhtar:2017ngi}, or from flavor composition of
extra-galactic neutrinos~\cite{Bustamante:2018mzu}. All of them,
however, largely depend on the assumptions made, and can be evaded in
specific scenarios.  In addition bounds from production of the light
$Z'$ in meson decays, neutrinoless $\beta\beta$ decay, or neutrino
annihilation in Big-Bang Nucleosynthesis (BBN) and supernovae are
relevant for $Z'$ with lower masses than those shown in the figure
(see, for example, Ref.~\cite{Dror:2020fbh}).

The bounds imposed by gravitational fifth force searches shown in
Fig.~\ref{fig:lowmass} were obtained by rescaling the results shown in
Ref.~\cite{Salumbides:2013dua} (which, in turn, were recasted from
Ref.~\cite{Adelberger:2009zz}). Limits from equivalence principle
tests are obtained rescaling the results from
Ref.~\cite{Schlamminger:2007ht}.  Details of the rescaling of the
published bounds applied for the specific models can be found in the
Appendix.  It is important to notice that these exclusion regions
obtained by recasting the boundaries of the published regions may not
correspond to the statistical condition we employed,
Eq.~\eqref{eq:dchibound}. So the comparison has to be taken with a
pinch of salt.  Still, from the figures we see that, generically, for
all models shown the oscillation analysis improves over the existing
bounds for $Z'$ lighter than $\sim 10^{-11}$~eV.

\begin{pagefigure}\centering
  \includegraphics[height=0.80\textwidth,angle=90]{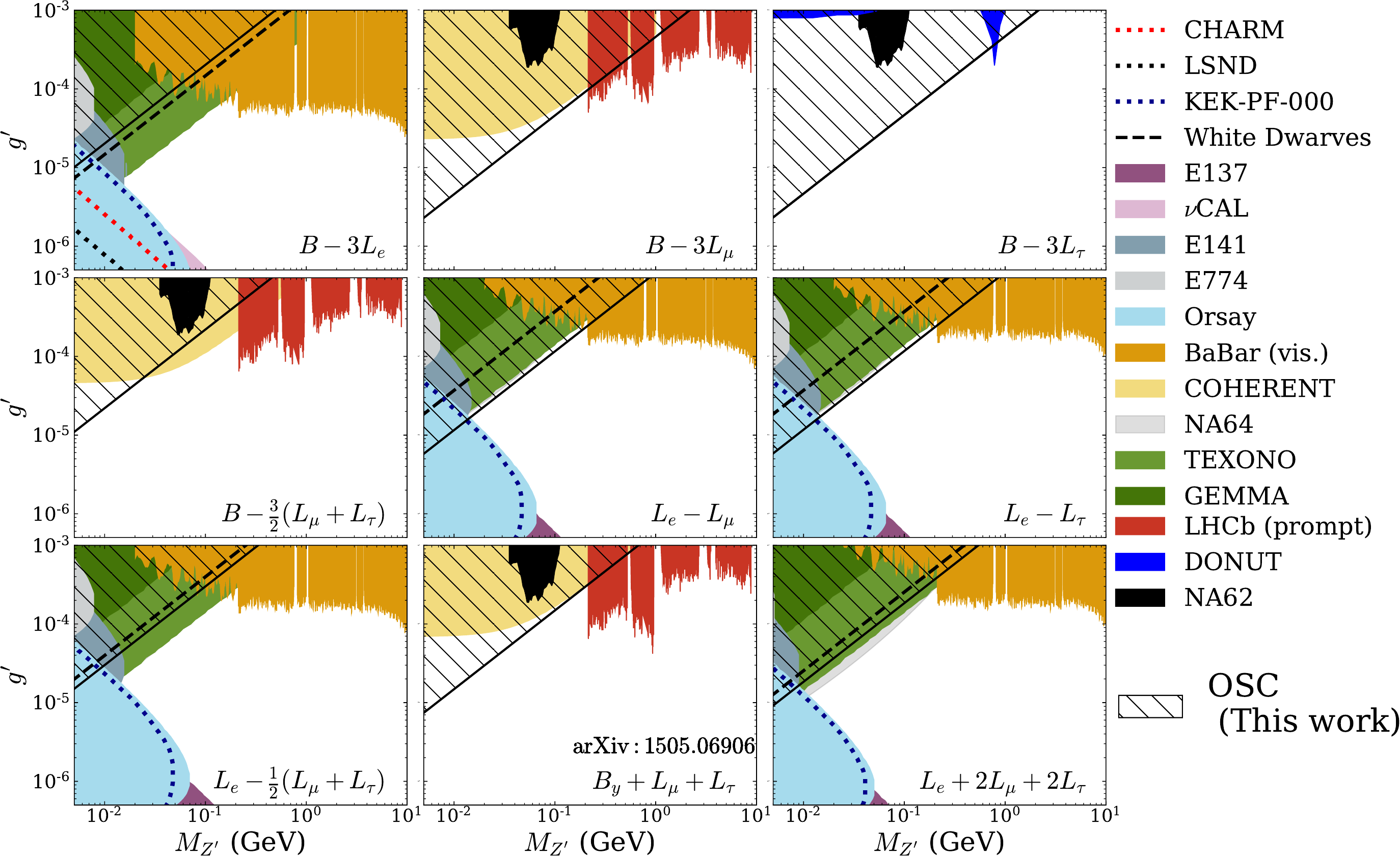}
  \caption{Values of $g'$ and $\text{5 MeV}\leq M_{Z'} \leq
    \text{10~GeV}$ for which a $U(1)'$ model coupled to the charges
    labeled in each panel gives a worse fit than standard oscillation
    by $2\sigma$, Eq.~\eqref{eq:dchibound} (hatched region).  We also
    show the bounds on these models imposed by a compilation of
    experiments as labeled in the figure.  In the window corresponding
    to the model in Ref.~\recite{Farzan:2015doa} we have defined $B_y
    \equiv B_1 - y B_2 - (3-y)B_3$.  See text for details.}
\label{fig:highmass}
\end{pagefigure}

Conversely the results shown in Fig.~\ref{fig:highmass} for $M_{Z'}$
in the $\mathcal{O}(\text{MeV--GeV})$ range correspond to $U(1)'$
effects in oscillation experiments always in the NSI domain.  In this
case, the analysis of oscillation data results in a bound on $g'$
which scales as the inverse of the mediator mass with coefficients
which we list in table~\ref{tab:boundsNSI}.  For the sake of
comparison, we also show in Fig.~\ref{fig:highmass} a compilation of
the most relevant experimental bounds on these $U(1)'$ models.  These
include constraints from electron and proton fixed target experiments,
neutrino electron elastic scattering, coherent neutrino nucleus
elastic scattering, white dwarf cooling and collider constraints.  Let
us point out that we have neglected kinetic mixing so far because it
does not affect the oscillation bounds as previously discussed.  But
it should be kept in mind that the presence of kinetic mixing could
either strengthen or weaken the bounds reported from other
experiments.  Appendix A contains all relevant details on the
derivation of these bounds.

As seen in Fig.~\ref{fig:highmass} for several models there are values
of $g'$ and $M_{Z'}$ which are only constrained by the oscillation
analysis.  This is particularly the case for $U(1)'$ coupled to $B -
3L_\tau$. We have found no competitive bound from other experiments in
the shown window with the only exception of the constraint reported in
Ref.~\cite{Kling:2020iar} using data from DONUT~\cite{Kodama:2007aa}
and from NA62 (for bounds from other experiments relevant for larger
couplings see for example Refs.~\cite{Farzan:2016wym,
  Heeck:2018nzc}).\footnote{Potentially, for the model in
Ref.~\cite{Farzan:2015doa}, coupled to $B_1 - y B_2 - (3-y) B_3 +
L_\mu + L_\tau$ there are additional strong bounds associated to quark
mixing effects due to the non-conservation of the $U(1)'$ current by
the SM particles (see Refs.~\cite{Dror:2017ehi, Dror:2017nsg,
  Dror:2018wfl}).  According to Ref.~\cite{Farzan:2015doa} such bounds
could be mitigated for a particular choice of $y$ and with the
inclusion of a $U(1)'$-charged scalar sector. But strictly speaking a
careful evaluation of these bounds has not been presented in the
literature, and it is beyond the scope of this paper.}

\subsection{Models for LMA-D}
\label{sec:dark}

A subset of models can lead to an allowed region in the DARK sector
with LMA-D within 4 units of $\chi^2$ with respect to the standard OSC
solution,
\begin{equation}
  \Delta\chi^2_\text{DARK,DOM}(g',M_{Z'}) < 4\,.
  \label{eq:dchilmad}
\end{equation}
As discussed in Sec.~\ref{sec:LR} this can only happen in the NSI
domain. In our survey we have found that 4.8\% of the set of charges
studied can have a best fit in LMA-D, and 5.2\% lead to LMA-D
verifying Eq.~\eqref{eq:dchilmad}. However none of these set of
charges correspond to the subclass of anomaly free vector $Z'$ models
with gauging of SM global symmetries with SM plus right-handed
neutrinos matter content (Eq.~\eqref{eq:anomod}).

In particular for the first seven set of models in
table~\ref{tab:models} the LMA-D solutions lies at more than $5\sigma$
from the standard oscillation fit. On the contrary the models in the
last two lines yield a viable LMA-D solution The first one was
proposed in Ref.~\cite{Farzan:2015doa} precisely as a viable model for
LMA-D.  Indeed in this case we find
$(\Delta\chi^2_\text{DARK,NSI})_\text{min} = 1.2$, which is within 4
units of $\chi^2$ from the pure oscillation result but the best fit
for this model charges still lies within the LIGHT sector. We also
show the results for a $U(1)'$ model coupled to $L_e + 2L_\mu +
2L_\tau$ for which we find that the best fit is LMA-D with
$(\Delta\chi^2_\text{DARK,NSI})_\text{min} = -1.3$.

\begin{figure}\centering
  \includegraphics[width=\textwidth]{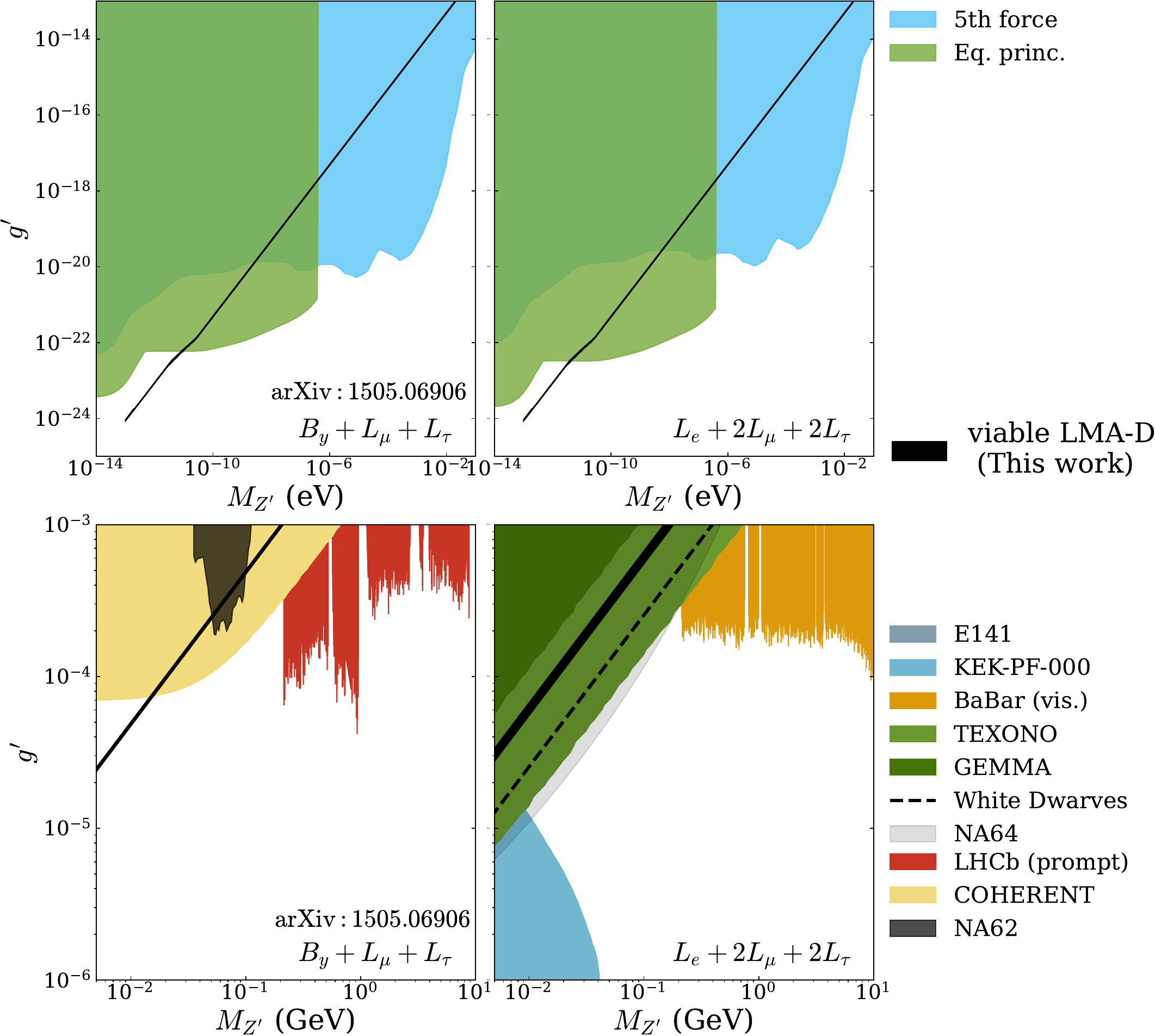}
  \caption{Range of $g'$ and $M_{Z'}$ for which the global analysis of
    oscillation data can be consistently described within the LMA-D
    solution, Eq.~\eqref{eq:dchilmad}, (black band) for two viable
    models.  We also show the bounds on these models imposed by a
    compilation of experiments as labeled in the figure.  In the
    window corresponding to the model in Ref.~\recite{Farzan:2015doa}
    we have defined $B_y \equiv B_1 - y B_2 - (3-y)B_3$.  See text for
    details.}
  \label{fig:dark}
\end{figure}

In Fig.~\ref{fig:dark} we plot as black bands the range of coupling
and masses for these two set of charges verifying the
condition~\eqref{eq:dchilmad}, together with the compilation of
relevant bounds from other experiments.

As seen in the figure, for the model in Ref.~\cite{Farzan:2015doa},
there are solutions for the $Z'$ coupling and mass for which LMA-D is
allowed without conflict with bounds from other experiments both for
very light mediators as well as for an $\mathcal{O}(\text{MeV--GeV})$
$Z'$.  More quantitatively for ultra light mediator part of the LMA-D
allowed parameter space for the model in Ref.~\cite{Farzan:2015doa} is
in conflict with the bounds from fifth force tests which impose the
stronger constraints for this model in this regime. But the LMA-D is
still a viable solution for
\begin{equation}
  10^{-13}~\text{eV} \leq M_{Z'} \leq 7\times 10^{-11}~\text{eV}
  \label{eq:lmad1}
\end{equation}
with couplings in a very narrow band and seen in the figure, for
example
\begin{equation}
  g'=(9.1\pm 0.2)\times 10^{-25}
  \quad\text{for}\quad
  M_{Z'} = 10^{-13}~\text{eV}
\end{equation}
and
\begin{equation}
  g'=(5.1\pm 0.2)\times 10^{-23}
  \quad\text{for}\quad
  M_{Z'} = 7\times 10^{-11}~\text{eV} \,.
\end{equation}
In addition LMA-D is also a viable solution for this model with
\begin{equation}
  5~\text{MeV} \leq M_{Z'} \leq 20~\text{MeV}
  \quad\text{with}\quad
  \frac{g'}{M_{Z'}} = \frac{(4.85\pm 0.15)\times 10^{-5}}{\text{MeV}}
\end{equation}
On the contrary, as seen in the figure, the model with charge
$L_e+2L_\mu+2L_\tau$ can only provide a viable LMA-D solution without
conflict with bounds from other experiments for and ultra light
mediator with $M_{Z'}\lesssim 5\times 10^{-12}$~eV.

\section{Conclusions}
\label{sec:conclu}

In this work we have performed dedicated global analysis of
oscillation data in the framework of lepton flavor dependent $U(1)'$
interactions which affect the neutrino evolution in matter.  The
analysis is performed for interaction lengths ranging from larger than
the Sun radius (covering what we label as LRI domain) to effective
contact neutrino interactions (NSI domain). We survey $\sim 10000$ set
of models characterized by the charges of the first generation charged
fermions and the three flavor neutrinos. We find that
\begin{itemize}
\item In the LIGHT sector of the oscillation parameter space the
  introduction of new interactions does not lead to a significantly
  better fit of the oscillation data compared to standard
  oscillations, irrespective of the $U(1)'$ coupling in either NSI or
  LRI domains. Thus for all cases the analysis of oscillation data in
  the LIGHT sector results in excluded ranges of $g'$ and $M_{Z'}$.

\item The excluded ranges for the $Z'$ coupling and mass for the
  models with couplings listed in table~\ref{tab:models} are shown in
  Figs.~\ref{fig:lowmass} and~\ref{fig:highmass} for $M_{Z'}$ below
  1~eV and in the $\mathcal{O}(\text{MeV--GeV})$ range, respectively
  -- \textit{i.e.}, below and above the window strongly disfavored by
  the cosmological bound on $\Delta N_\text{eff}$.

\item In the regime of ultra-light mediators, for all models shown,
  the oscillation analysis improves over the bounds from tests of
  fifth forces and of violation of equivalence principle and for $Z'$
  lighter than $\sim 10^{-11}$~eV.

\item For mediators in the $\mathcal{O}(\text{MeV--GeV})$ range we
  list in table~\ref{tab:boundsNSI} the derived constraints on $g'$
  versus $M_{Z'}$. We find that for several of the considered models
  there are values of $g'$ and $M_{Z'}$ for which the oscillation
  analysis provides constraints extending beyond those from other
  experiments.

\item In what respects to LMA-D we find that it cannot be realized in
  the LRI domain. In the NSI domain we have found that 4.8\% of the
  set of charges studied can have a best fit in LMA-D, and 5.2\% lead
  to LMA-D as a valid solution within 4 units of $\Delta\chi^2$ of
  standard oscillations. None of these set of charges correspond to an
  anomaly free model based on gauging SM global symmetries with SM
  plus right-handed neutrinos matter content (Eq.~\eqref{eq:anomod}).
  So, generically, $Z'$ models for LMA-D with gauged SM global
  symmetries require additional states for anomaly cancellation.
\end{itemize}

\subsection*{Acknowledgement}

We want to thank Renata Zukanovich for discussions and her
participation in the early stages of this work.  The authors also
acknowledge use of the HPC facilities at the IFT (Hydra cluster) This
work was supported by the Spanish grants FPA2016-76005-C2-1-P,
FPA2016-78645-P, PID2019-105614GB-C21 and PID2019-110058GB-C21, by EU
Network FP10 ITN HIDDEN (H2020-MSCA-ITN-2019-860881), by USA-NSF grant
PHY-1915093, by AGAUR (Generalitat de Catalunya) grant
2017-SGR-929. The authors acknowledge the support of the Spanish
Agencia Estatal de Investigación through the grant ``IFT Centro de
Excelencia Severo Ochoa SEV-2016-0597''. PC acknowledges funding from
the Spanish MICINN through the ``Ramón y Cajal'' program under grant
RYC2018-024240-I.

\appendix
\section{Bounds from non-oscillation experiments}
\label{sec:appendix}

Bounds on dark photons or light $Z'$ bosons can be set using a variety
of laboratory experiments, collider searches and cosmological and
astrophysical probes. While in this work we focus on the most relevant
ones for the range of masses under consideration, we refer the
interested reader to Refs.~\cite{Ilten:2018crw, Fabbrichesi:2020wbt,
  Bauer:2018onh, Wise:2018rnb, Jaeckel:2010ni, Essig:2013lka,
  Harnik:2012ni} for a selection of works which include a systematic
compilation of bounds.  In this appendix we summarize the main details
of the bounds shown by the colored regions in Figs.~\ref{fig:lowmass}
and~\ref{fig:highmass}, as well as the procedure used to rescale them
for the different models shown in each panel.

\subsection*{Bounds from searches for gravitational fifth forces}

Generically gravitational fifth force experiments look for deviation
from the standard Newton potential ($\propto 1/r$) between two
objects.

We have taken the results from Ref.~\cite{Salumbides:2013dua} (which,
in turn, were recasted from Fig.~10 in Ref.~\cite{Adelberger:2009zz})
where they present the constraints on the coupling $\alpha_5$ versus
the interaction length $\lambda$ (or, equivalently, versus $m_5 =
\frac{1}{\lambda}$) defined as the constant entering the potential of
the fifth force
\begin{equation}
  \label{eq:V5}
  V_5(r) = \alpha_5\, \mathcal{N}_1\, \mathcal{N}_2\,
  \frac{e^{-r/\lambda}}{r}
\end{equation}
where $\mathcal{N}_{1,2}$ is the total charge of each object, which
they take as the total number of baryons.

In particular from Fig.~3 in~\cite{Salumbides:2013dua} we read their
boundary curve
\begin{equation}
  \left( \frac{\alpha^\text{max}_5}{\alpha_{em}}\,,\, m_5 \right) \,,
\end{equation}
where $\alpha_{em}$ is the SM fine structure constant.  In order to
rescale it to the different $U(1)'$ models we notice that the
corresponding potential for the $U(1)'$ interaction for the same
objects is
\begin{equation}
  V'(r) = C_1 C_2 \frac{{g'}^2}{4\pi} \frac{e^{-r/\lambda'}}{r} \,,
  \label{eq:potZ'}
\end{equation}
where $C_i$ is the total $Z'$ charge of object $i$ and $\lambda'$ is
the $U(1)'$ interaction length
\begin{equation}
  \lambda' = \frac{1}{M_{Z'}} \,,
  \quad
  C_i \equiv \frac{\mathcal{N}_i} {A_i} c_i
  \equiv \frac{\mathcal{N}_i}{A_i}
  \left[Z_i(a_e + 2a_u + a_d)+ (A_i - Z_i)(a_u + 2a_d) \right] \,,
\end{equation}
where $Z_i$ and $A_i$ are the atomic number and mass number of the
material of which the object $i$ is made, so $c_i$ is the charge under
$Z'$ for each atom of the material. Assuming that the number of
protons and neutrons in the material are not very different (that is,
$A_i / Z_i \sim 2$), we get
\begin{equation}
  V'(r)= V_5(r) \times \frac{g'^2}{4\pi\alpha_5} \frac{(a_e+3a_u+3a_d)^2}{4}
\end{equation}
with $m_5 = M_{Z'}$. So the boundary in the $g'$ vs $M_{Z'}$ plane
will be
\begin{equation}
  \big( g'_\text{max}, M_{Z'} \big) =
  \left(\frac{2}{a_e+3a_u+3a_d} \sqrt{4\pi\alpha_{em}} \times
  \sqrt{\frac{\alpha^\text{max}_5}{\alpha_{em}}} \,,\, m_5\right) \,.
\end{equation}

\subsection*{Bounds from searches for violation of the equivalence principle}

Similarly as in the case of fifth force searches, this bound comes
from precise measurements of the gravitational potential between two
objects. However, in this case one tests the differences of the
potential for the same total mass of two different test materials,
using a pendulum which is attracted by the same mass. The bounds are
taken from Ref.~\cite{Schlamminger:2007ht}, where they define the
potential due to the new force as:
\begin{equation}
  V_G(r) = \alpha\, G  \frac{m_t m_s}{r}
  \hat{\mathcal{N}}_t \hat{\mathcal{N}}_s e^{- r / \lambda } \,.
\end{equation}
where $G$ is the gravitational constant, $\hat{\mathcal{N}}_i$ refers
to the new interaction charge per mass unit, and the subindices $t$
and $s$ stand for test (or pendulum) and source masses,
respectively. In Ref.~\cite{Schlamminger:2007ht} they assume that the
new interaction couples to the number of baryons. Thus, they use
beryllium and titanium as test materials, chosen to maximize the
difference in baryon number per unit mass. To be specific, we use
$\hat{\mathcal{N}}_\text{Be} = 0.99868$ and
$\hat{\mathcal{N}}_\text{Ti} = 1.001077$, as in
Ref.~\cite{Schlamminger:2007ht}.

Noting that the total potential will be the sum of the standard
gravitational potential plus the contribution from the new
interaction, the authors of Ref.~\cite{Schlamminger:2007ht} put a
bound on
\begin{equation}
  \eta = 2\frac{V_\text{Be} - V_\text{Ti}}{V_\text{Be} + V_\text{Ti}}
  \sim \frac{\Delta V_G(r)}{\frac{G m_s m_t}{r}}
  = \alpha(\hat{\mathcal{N}}_\text{Ti} - \hat{\mathcal{N}}_\text{Be})
  \hat{\mathcal{N}_s} e^{-r / \lambda}
\end{equation}
which is used to set a constraint on $\alpha$ as a function of
$\lambda$,
\begin{equation}
  \big( \alpha^\text{max} \,,\, \lambda \big) \,.
\end{equation}
This can be used to set a bound on $\Delta V'(r)$ for a general model,
noting that
\begin{equation}
  \begin{split}
    \Delta  V'(r)
    &= \frac{{g'}^2}{4\pi} \frac{e^{-r M_{Z'}}}{r} \,C_s
    (C_\text{Be} - C_\text{Ti})
    \\
    &= \Delta V_G(r) \times \frac{g'^2/4\pi} {\alpha G\cdot u^2}
    \times \frac{c_\text{Ti}/A_\text{Ti} - c_\text{Be}/A_\text{Be}}{\hat{\mathcal{N}}_\text{Ti}-\hat{\mathcal{N}}_\text{Be}}
    \times \frac{c_s/A_s}{\hat{\mathcal{N}}_s}
  \end{split}
\end{equation}
where $u$ stands for the atomic mass unit in GeV ($u\simeq
0.931$~GeV).
So the boundary in the $g'$ vs $M_{Z'}$ plane will be
\begin{equation}
  \big(g'_\text{max}\,,\, M_{Z'} \big)
  = \left(\sqrt{\frac{\hat{\mathcal{N}}_\text{Ti}-\hat{\mathcal{N}}_\text{Be}}{c_\text{Ti}/A_\text{Ti} - c_\text{Be}/A_\text{Be}}
    \times \frac{\hat{\mathcal{N}}_s}{c_s/A_s} \, 4\pi\, G\, u^2}
  \times \sqrt{\alpha^\text{max}} \,,\, \frac{1}{\lambda} \right) .
\end{equation}

\subsection*{Bounds from white-dwarf cooling}

We based the bounds shown in Figs.~\ref{fig:highmass}
and~\ref{fig:dark} on the study presented in
Ref.~\cite{Dreiner:2013tja}. There, upper bounds on new interactions
are imposed on the basis that the energy losses from plasmon decays
into particles that escape the star is not larger than the energy
losses due to neutrino emission in the SM.

In the $U(1)'$ scenarios here considered the minimum new contribution
is due to $Z'$ mediated decays into neutrinos
\begin{equation}
  \label{eq:plasmon-wd}
  \Gamma_{\text{plasmon} \to \nu \bar\nu, Z'}^s \lesssim
  \Gamma_{\text{plasmon} \to \nu \bar\nu, \text{SM}}^s =
  \frac{C_{e,V}^2 G_F^2}{48\pi^2 \alpha_{em}}\frac{Z_s \pi_s^3}{\omega_s}
\end{equation}
where $C_{e,V} $ is the vector coupling to the electron current in the
SM, $Z_s$ is the plasmon wavefunction renormalization and $\pi_s$ is
the effective plasmon mass which enters in the dispersion relation
$\omega^2_s - k^2 = \pi_s(\omega_s, k)$. Here, $s=T,L$ refers to the
plasmon polarization (transverse or longitudinal).

Under the assumption that the mass of the $Z'$ is much larger than the
frequency of the plasmon we can write its rate into neutrinos of a
given flavor $\beta$ due to the new interactions as
\begin{equation}
  \Gamma_{\text{plasmon} \to \nu_\beta \bar\nu_\beta, Z'}^s =
  \frac{1}{3}\frac{{g'}^4}{M_{Z'}^4}
  \frac{(a_e\,b_\beta)^2}{48\pi^2 \alpha_{em}}
  \frac{Z_s \pi_s^3}{\omega_s}
\end{equation}
And the upper bound obtained in Ref.~\cite{Dreiner:2013tja} translates
into:
\begin{equation}
  \sqrt{\sum_\beta\frac{(a_e\, b_\beta)^2}{3}}
  \cdot \frac{{g'}^2}{M_{Z'}^2}
  \leq C_{e,V}\, G_F = 1.12\times 10^{-5}~\text{GeV}^{-2} \,.
\end{equation}

\subsection*{Bounds from coherent neutrino-nucleus scattering }

We have performed our own reanalysis of coherent neutrino-nucleus
scattering data using the time and energy information from COHERENT
experiment~\cite{Akimov:2017ade, Akimov:2018vzs} on CsI based on our
recent analysis in Ref.~\cite{Coloma:2019mbs} performed for NSI with a
variety of nuclear form factors, quenching factors and parametrization
of the background.  In particular the results shown in
Fig.~\ref{fig:highmass} correspond to the analysis performed using the
quenching factor obtained with the fit to the calibration data of the
Duke (TUNL) group~\cite{Akimov:2017ade} together with our data driven
reevaluation of the steady-state background (see
Ref.~\cite{Coloma:2019mbs} for details).

Model predictions are obtained exactly as in~\cite{Coloma:2019mbs},
but replacing the cross section for coherent scattering in the
presence of NSI with that induced by the $Z'$. In particular if we
define
\begin{equation}
  \label{eq:eps-prime}
  \varepsilon'(Q^2, g', M_{Z'})
  \equiv \frac{1}{\sqrt{2} G_F}\frac{{g'}^2}{M_{Z'}^2 + Q^2}
\end{equation}
one can use the same cross section expression for NSI simply
replacing:
\begin{equation}
  \label{eq:NSI-repl}
  \varepsilon_{\alpha\beta}^{q,V} \to \delta_{\alpha\beta}\, a_q\, b_{\alpha}\,
  \varepsilon'(Q^2, g_X, M_{Z'})
\end{equation}
where $Q^2 = 2 M T$ is the momentum transfer. In this case, the
differential cross section for coherent scattering of a neutrino with
flavor $\alpha$ reads:
\begin{equation}
  \frac{d\sigma_\alpha}{dT} = \frac{G_F^2}{2\pi}
  W_\alpha^2 (Q^2, g_X, M_{Z'}) F^2(Q^2) M
  \left( 2 - \frac{MT}{E_\nu^2} \right)
\end{equation}
where $M$ is the mass of the nucleus, $T$ is the nuclear recoil
energy, and $E_\nu$ is the incident neutrino energy. We have defined a
modified weak charge for the nucleus as:
\begin{equation}
  \begin{split}
    \label{eq:W_alpha}
    W_\alpha (Q^2, g_X, M_{Z'})
    & = Z\big[ g_p^V + (2 a_u + a_d)\, b_\alpha\,
      \varepsilon'(Q^2, g_X, M_{Z'}) \big] +
    \\
    & + N\big[ g_n^V + (a_u + 2 a_d)\, b_\alpha\,
      \varepsilon'(Q^2, g_X, M_{Z'}) \big] \,.
  \end{split}
\end{equation}
Thus, unlike in the case of NSI, the weak charge now depends on the
momentum transferred in the process.

The construction of $\chi^2_\text{COH}(g', M_{Z'})$ is totally
analogous to that of $\chi^2_\text{COH} (\vec\Eps)$ in
Ref.~\cite{Coloma:2019mbs}.  In consistency with the condition impose
when deriving the oscillation bounds the COHERENT regions shown in
Fig.~\ref{fig:highmass} corresponds to the $Z'$ coupling and mass for
which the fit to COHERENT data is worse than the one obtained in the
SM by 4 units:
\begin{equation}
  \chi^2_\text{COH}(g', M_{Z'}) - \chi^2_\text{COH,SM} > 4.
\end{equation}

Although it is difficult to compare to other works in the literature
due to the very different implementation of systematics, quenching
factor, form factor and background treatment, we find that our results
are roughly consistent with those presented in
Refs.~\cite{Cadeddu:2020nbr, Papoulias:2019txv, Denton:2018xmq}.

\subsection*{Bounds from measurements of neutrino scattering on electrons}

For neutrino-electron scattering experiments TEXONO and GEMMA we
performed our own analysis following the procedure in
Ref.~\cite{Lindner:2018kjo} which explicitly studied the bounds
imposed by those experiments in some $Z'$ models and provide all the
relevant cross section expressions. However, note that the current of
our Lagrangian differs by a factor of two with respect to the one in
Ref.~\cite{Lindner:2018kjo}. This has been accounted for in our
calculations.

For TEXONO we use the data from Fig.~16 in Ref.~\cite{Deniz:2009mu}.
This corresponds to 29882 (7369) kg-day of fiducial mass exposure
during Reactor ON (OFF), respectively. The adopted analysis window is
3--8 MeV, spread out uniformly over $N_\text{bin} = 10$ energy
bins. An overall normalization constant has been manually set to
reproduce the SM prediction shown in Fig 16 in~\cite{Deniz:2009mu}.
Cross section has been implemented following~\cite{Lindner:2018kjo}.
With this we get $\chi^2_\text{SM,TEXONO}= 8.75$.

For GEMMA we use the Phase I data shown in Fig.~8 of
Ref.~\cite{Beda:2010hk}.  This corresponds to about 5184 ON-hours and
1853 OFF-hours of active time.  Data was taken for a detector mass of
1.5~kg.  The energy window used is $0.015~\text{MeV} \leq E_\nu \leq
8.0~\text{MeV}$. With this procedure we get $\chi^2_\text{SM,GEMMA} =
26.33$, for 39 energy bins.

For both TEXONO and GEMMA we draw the contours with the equivalent
condition used for the oscillation analysis so for a given set of
model charges the contours are defined by $\chi^2(g',M_{Z'}) -
\chi^2_\text{SM} = 4$.

Additional bounds from neutrino scattering on electrons can be derived
from the analysis of LSND or CHARM II data (see, \textit{e.g.},
Ref.~\cite{Bilmis:2015lja}).  However, they are less constraining than
those obtained for TEXONO and GEMMA in the shown range of masses. They
can be stronger for masses above $\sim 50$~MeV but in such case they
are weaker than bounds from other experiments, in particular in $e^+
e^-$ collisions in BaBaR~\cite{Lees:2014xha} (see below). For this
reason they are not shown in the figures.

\subsection*{Fixed target experiments and colliders}

As mentioned in the text, the bounds from fixed target experiments and
colliders are taken directly from the literature. In particular, we
consider the following set of bounds:
\begin{itemize}
\item Electron beam dump experiments: we take these from the
  compilation in Ref.~\cite{Andreas:2012mt}, which were obtained using
  data from E137~\cite{Bjorken:1988as}, E141~\cite{Riordan:1987aw},
  E774~\cite{Bross:1989mp}, Orsay~\cite{Davier:1989wz}, and
  KEK-PF-000~\cite{Konaka:1986cb}.

\item Proton beam dump experiments: for LSND we use the results
  obtained in Ref.~\cite{Essig:2010gu} which used data from
  Ref.~\cite{Auerbach:2003fz} (see also Ref.~\cite{Batell:2009di});
  for CHARM~\cite{Bergsma:1985qz} we use the limit derived in
  Ref.~\cite{Gninenko:2012eq}; finally, for
  $\nu$-Cal~\cite{Blumlein:1990ay} we use the lmit derived in
  Ref.~\cite{Tsai:2019mtm} (see also Ref.~\cite{Blumlein:2011mv} for a
  similar analysis).

\item We also consider constraints from $Z'$ production in $e^+ e^-$
  collisions in BaBaR both in visible~\cite{Lees:2014xha} and
  invisible~\cite{Lees:2017lec} final states, as well as constraints
  from LHCb for $U(1)'$ decaying into $\mu^+\mu^-$, the most relevant
  ones being from prompt decay searches~\cite{Aaij:2019bvg,
    Ilten:2016tkc}.

\item Constraints on invisible $Z'$ decays in searches for $\pi^0\to
  \gamma\, Z'$ at NA62 experiment~\cite{CortinaGil:2019nuo} and in
  $e\, N\to e\, N\, Z'$ at NA64 experiment~\cite{Banerjee:2016tad,
    Banerjee:2017hhz, NA64:2019imj}.  There are bounds from NA64 for
  the $Z'$ decaying into $e^+e^-$~\cite{Banerjee:2019hmi}, but for the
  models in Fig.~\ref{fig:highmass} they are weaker than those from
  other experiments.
\end{itemize}

All the bounds mentioned above were obtained for a dark photon coupled
to the SM fermions via kinetic mixing. In order to recast these to
bounds on the different $U(1)'$ models considered we have used the
\texttt{darkcast} software~\cite{Ilten:2018crw}, which takes into
account the difference in production branching ratio and lifetime of
the new boson, as well as its decay into a given final state.  In
particular for invisible decays we include only the decays into
neutrinos.

In addition bounds on $U(1)'$ models coupled to charges including
$L_\mu$ can be constrained with data on production of $\mu^ + \mu^-$
in $\nu_\mu$ scattering off the Coulomb field of a nucleus (the so
called neutrino trident production) at CHARM-II~\cite{Geiregat:1990gz}
and the CCFR experiment~\cite{Mishra:1991bv}, see
Ref.~\cite{Altmannshofer:2014pba}. They can also be bounded with data
on $\mu^+\mu^-$ production in $e^+ e^-$ collision at
BaBar~\cite{TheBABAR:2016rlg} and Belle~\cite{Adachi:2019otg}.  We
have verified that all these bounds are always weaker than those
imposed by either coherent neutrino-nucleus scattering, or LHCb for
the same models. For this reason they are not shown in the figures.

\subsection*{Astrophysical and cosmological bounds for $M_{Z'}\gtrsim \mathcal{O}(\text{MeV})$}

The impact of light $Z'$ on astrophysical and cosmological observables
can also be used to set strong constraints on these models.

For example, non-standard cooling mechanisms in the Sun, other stars
or supernovae from $Z'$ emission can be used to set very tight
constraints on these models. Solar and stellar constraints are only
relevant in the mass window $ \textrm{eV} \lesssim M_{Z'} \lesssim
100~\textrm{eV}$ (see, \textit{e.g.}, Refs.~\cite{Harnik:2012ni,
  Wise:2018rnb, Jaeckel:2010ni, Davidson:2000hf}) and therefore will
not be considered here. More relevant are the SN constraints, which
also apply to masses above a few MeV and therefore would be relevant
in the high mass window considered in this work. However, while these
have been derived in the dark photon scenario, the production
mechanisms would be significantly affected in the $Z'$ case. As an
example, in Ref.~\cite{Croon:2020lrf} a specific analysis carried out
for supernovae emission for the $L_\mu - L_\tau$ and $B-L$ models
showed large variations in the results with respect to the region
constrained in the dark photon case~\cite{Raffelt:2006cw,
  Chang:2016ntp}. A dedicated analysis would be required to adapt
these bounds to models with arbitrary charges.

Finally additional constraints can be derived from cosmological
observations, from the energy injection of the $Z' $ onto $e^+ e^-$ in
the early Universe, which would be applicable in the mass region above
1~MeV (see, \textit{e.g.}, Fig.~6 in~\cite{Coffey:2020oir}). However,
again in this case it is uncertain how to recast these bounds to a
general $U(1)'$ model with arbitrary charges, and therefore we have
not included them here.

\bibliographystyle{JHEPmod}
\bibliography{references}

\end{document}